\documentclass[10pt,aps,prb,twocolumn,floatfix,longbibliography]{revtex4-2}
\usepackage{amsmath,slashed,amsfonts,graphicx,verbatim}
\usepackage[colorlinks=true, linkcolor=blue, citecolor=blue, urlcolor=blue]{hyperref}
\usepackage{xcolor}
\usepackage{soul}

\begin{document}
\title{Interaction robustness of the chiral anomaly in Weyl semimetals and Luttinger liquids from a mixed anomaly approach}
\author{Shuyang Wang}
\author{Jay D. Sau}
\affiliation{Condensed Matter Theory Center and Joint Quantum Institute, Department of Physics, University of Maryland, College Park, Maryland 20742, USA}
\begin{abstract}
 The chiral anomaly is one of the robust quantum effects in relativistic field theories with a chiral symmetry where charges in chiral sectors appear to be separately conserved. The chiral anomaly, which is often associated with a renormalization-invariant topological term, is a violation of this conservation law due to quantum effects. Such anomalies manifest in Weyl materials as an electromagnetic field-induced transfer of charge between Fermi pockets. However, the emergent nature of the conservation of chiral charge leads to manifestations of the chiral anomaly response that depend on the details of the system such as the strength of interactions. In this paper, we apply an approach in which the chiral symmetry in solid materials is viewed as a so-called emanant symmetry from the underlying space translation symmetry. The chiral anomaly in this case is replaced by a mixed anomaly between the charge $U(1)$ and the space translation, while the chiral charge can be defined as proportional to the total momentum. We show that the chiral anomaly associated with this chiral charge is unrenormalized by interactions, in contrast to other chiral charges in $(1+1)D$ whose renormalization is regularization dependent. In $(3+1)D$ Weyl systems, this chiral anomaly is equivalent to the charge transferred between Fermi surfaces, which can be measured through changes in Fermi-surface-enclosed volume. We propose a pump-probe technique to measure this.
\end{abstract}

\maketitle
\section{Introduction}
The discovery of Weyl and Dirac materials \cite{Wan2011,Novoselov2004,Armitage2018} with low energy quasiparticles that are described as a $(3+1)$D Dirac equation has motivated
The study of condensed matter analogs of the chiral anomaly of relativistic fermions in high-energy physics.
Fundamentally, the chiral anomaly in high-energy physics~\cite{adler1969axial,bell1969nuovo} arises from the fact that the electromagnetic response of the regularized charge of $\mathrm{QED}_{2n}$ with a single Weyl fermion breaks gauge invariance (i.e. charge conservation) despite the apparent gauge invariance of the action of the Weyl fermion. The gauge invariance can be restored by introducing a second Weyl fermion, which then leads to a chiral symmetry. 
The chiral symmetry is a $U(1)$ gauge symmetry similar to the one associated with charge conservation except that it assigns opposite so-called chiral charges to the different Weyl fermions referred to as chiral sectors. 
The response of such a pair of Weyl fermions can be gauge invariant in addition to being Lorentz invariant, although it 
necessarily breaks conservation of the chiral charge current $j^{5\mu}$~\cite{adler1969axial,bell1969nuovo}.
The equation for violation of chiral charge current conservation is the so-called chiral anomaly equation~\cite{Peskin2015}:
\begin{equation}\label{anomaly}
\partial_\mu j^{5\mu}=(-1)^{n+1}\frac{2e^n}{n!(4\pi)^n}\epsilon^{\mu_1\mu_2\cdots\mu_{2n}}F_{\mu_1\mu_2}\cdots F_{\mu_{2n-1}\mu_{2n}}.
\end{equation}

The chiral symmetry discussed above, which corresponds to the conservation of electron density at each Fermi pocket, is not a microscopic symmetry of the Weyl and/or Dirac systems and instead is an emergent IR symmetry~\cite{else2021Non-Fermi} that appears after projecting excitations to the low-energy limit. The essential dynamics underlying the chiral anomaly for $n=2$ is the one in parallel electric and magnetic fields~\cite{Nielsen1983} and can be seen to arise as a consequence of the Berry curvature near the Weyl point.
We can quantify the chiral anomaly by projecting the quasiparticles to a shell around the Fermi surface, which is analogous to the regularization discussed in the previous paragraph.
This procedure when applied to non-interacting systems predicts a counter-intuitive transfer of charge between the difference Fermi surfaces~\cite{burkov2015chiral,Armitage2018}
in response to an electromagnetic field.

The above formulation of the chiral anomaly in solid state systems applies only to non-interacting systems, which leads to the question of whether the chiral anomaly (i.e. Eq.~\ref{anomaly}) 
is affected by interactions. 
In fact, the chiral anomaly in $(3+1)D$ relativistic quantum field theory (i.e. $\mathrm{QED}_4$) written in Eq.~\ref{anomaly} is found not to be affected by the addition of interactions in the sense
that it does not receive radiative corrections~\cite{Adler1969}.
This robustness can be understood in situations where the anomaly can be constrained by a topological term in an action on a higher-dimensional space~\cite{Peskin2015,wen2004quantum,zhang2001four}.
On the other hand, the chiral anomaly in the $(1+1)D$ Thirring model, another relativistic quantum field theory, is found to be renormalized by interactions~\cite{Hagen1967, Georgi1971, Shei1972} 
according to the equation
\begin{equation}\label{geo}
\partial_\mu j^{5\mu}=\left(\frac{1}{1+\lambda/\pi}\right)\frac{e}{\pi}F_{01},
\end{equation}
where $\lambda$ is the Thirring interaction strength.

Given that solid state systems can support a variety of interaction terms, it is natural to discuss modifications to the chiral anomaly by interactions~\cite{Avdoshkin2020,Rylands2021,parhizkar2023path}. For example, the inclusion of Umklapp scattering explicitly breaks the chiral symmetry even in the IR limit, invalidating the notion of a chiral anomaly. In contrast to interaction robustness of the chiral anomaly, the disorder robustness of the chiral anomaly for intra-valley 
scattering 
has been established both semiclassically~\cite{Son2013} and fully quantum mechanically using the non-linear sigma model~\cite{lee2018chiral}.
All of these discussions of the chiral anomaly in solid-state systems are complicated by the fact that the chiral charge is only defined in the IR limit and therefore difficult to measure as a microscopic variable. 
As a result, much of the experimental discussion of the chiral anomaly has focused on manifestations such as the chiral magnetic effect~\cite{Armitage2018} and the negative magnetoresistance~\cite{Son2013}, which, although relatively convenient to measure, are known to contain contributions from other mechanisms~\cite{goswami2015axial}.
In fact, studies of the effects of interactions on the current response, which are signatures of the anomaly, such as the chiral photocurrent~\cite{Avdoshkin2020} and the chiral current~\cite{Rylands2021} have shown a renormalization by the interaction of these response functions similar to Eq.~\ref{geo}.
Interestingly, the detection of the chiral anomaly via non-local transport~\cite{parameswaran2014probing} provides a way to detect spatial variations of the chiral anomaly coefficient in Eq.~\ref{anomaly} without being directly sensitive to the coefficient itself. 
On the other hand, it is possible to find an interaction-robust chiral anomaly in a slab of a four-dimensional second Chern insulator~\cite{avron1988topological,zhang2001four} viewed as a three-dimensional lattice system. The difference in charge between surface states of such a Chern insulator defines a chiral charge, while the second Chern-Simons form of the four-dimensional quantum Hall effect defines the quantized chiral anomaly.
These solid-state examples together with the two relativistic theories (Eq. 1 and 2) suggest that the magnitude of the chiral anomaly in the absence of Lorentz invariance can depend on both the details of the interaction Hamiltonian and the regularization of chiral charge.
To illustrate the latter, as we will elaborate later, an interacting Weyl semimetal can be described as a topological Fermi liquid\cite{Haldane2004} of non-interacting quasiparticles. In this limit, the chiral anomaly with the chiral charge defined in terms of quasiparticles becomes topological and is not renormalized.


The goal of this work is to characterize the effect of interactions on the chiral anomaly equation (i.e. Eq.~\ref{anomaly}) in several models of interaction as well as the expectation values of chiral charge and elaborate on a few cases where the anomaly can be defined in a topologically robust way. For the latter example, we will focus on a continuum model of an interacting Weyl semimetal where the chiral symmetry can be viewed as an emanant symmetry~\cite{Cheng_2023} emanating from the UV space translation symmetry.
The chiral anomaly is then viewed as a mixed anomaly between the charge $U(1)$ and translation-symmetry~\cite{Oshikawa2000,Cho2017,wang2021emergent}, which is topological and therefore not affected by interactions. 

To achieve this goal, we will start by calculating the chiral anomaly in several examples with interaction. Specifically, in Sec.~\ref{4} by comparing the results for the anomaly in $1+1$D systems i.e. Eq.~\ref{geo} using relativistic regularizations~\cite{Georgi1971} and non-relativistic bosonization of the Tomonaga-Luttinger (TL) model we will be able to illustrate the regularization dependence of the anomaly. We will argue how regularization can ultimately be viewed as a redefinition of chiral charge. We also discuss how refermionization~\cite{von1998bosonization} of the Sine-Gordon model might be used to define a relativistic chiral anomaly in non-relativistic systems. In Sec.~\ref{3}, we use a continuum version of the mixed anomaly between $U(1)$ and translation-symmetry~\cite{Oshikawa2000,Cho2017} to define an equation that is closely analogous to the chiral anomaly equation. This mixed anomaly has previously been used~\cite{wang2021emergent} to predict the action of the microscopic chiral Landau level from the topological term associated with the emergent IR chiral symmetry. We find that this approach leads to a chiral charge that is based on momentum density and a corresponding anomaly equation that is not renormalized by interactions. 
This kind of chiral charge has been surprisingly used and measured in the $^3$He systems\cite{Bevan1997volovik}. 
We also explore an intriguing application of this momentum-based chiral charge conservation law in a junction configuration between an integer quantum Hall (IQH) edge (\(\nu = 1\)) and a fractional quantum Hall (FQH) edge (\(\nu = 1/3\)), as seen in a recent experiment \cite{Cohen2023}.

In Sec.~\ref{6}, we apply the mixed anomaly approach to define the chiral anomaly of Weyl semimetals
starting with the limit of weak magnetic fields and finite chemical potential where the topological Fermi liquid theory~\cite{Haldane2004} may be applied. 
The results apply to continuum approximations for time-reversal breaking Weyl materials~\cite{Burkov2018,Ahn2017}, 
which have some subtleties that are discussed later.
We find that the chiral anomaly in terms of the chiral charge defined as the difference of charge density between the Weyl points can be derived from the mixed anomaly generalized to $(3+1)$D\cite{wang2021emergent,hughes2024gapless}. This change in chiral charge is associated with a change in the Luttinger volume of each of the Fermi surfaces. We propose a pump-probe technique to directly measure the change in Luttinger volume that is robust to interactions. While this is more difficult to measure than the current response from the chiral magnetic effect, it does not depend on additional transport measurements and directly measures the transfer of charge between Weyl points, which is the unique feature of the anomaly.

\section{Regularization Dependence of the Chiral Anomaly in \texorpdfstring{$(1+1)$D}{(1+1)D}\label{4}}
The Tomonaga-Luttinger model~\cite{Giamarchi2007}, which describes low-energy excitations in $(1+1)$D systems
with fermionic excitations, has an emergent Lorentz symmetry. 
This emergent Lorentz symmetry motivates the study of the chiral anomaly using methods of high-energy physics such as Fujikawa's method, Pauli-Villars regularization, and Feynman diagrams\cite{Fujikawa1980,Adler1969,DIAZ1989}.
The action of the Luttinger model, $S_{LL}$, is written as $S_{LL}=S_{LL,0}+S_{LL,int}$ with the Fermi velocity ($v_F=1$)
where 
\begin{equation}{\label{lut}}
S_{LL,0}=\int d^2x \, i\bar{\Psi}(\slashed{\partial}+ie\slashed{A})\Psi,
\end{equation}
is the action associated with the non-interacting fermionic field $\Psi:=(\psi_R,\psi_L)^T$ composed of left and right movers with no spins,
and $A$ is the electromagnetic vector potential.
Furthermore, $\bar{\Psi}=\Psi^\dagger\gamma^0$, and the Feynman slash notation $\slashed{a}:=a_\mu\gamma^\mu$ is used. 
The action for the Luttinger model, $S_{LL}$, can also include a general local current-current interaction of the form\cite{Rylands2021}:
\begin{equation}{\label{lutint}}
S_{LL,int}=-\frac 12 \int d^2x\, \lambda^{(2)}_{\mu\nu} j_C^\mu j_C^\nu,
\end{equation}
where 
\begin{equation}\label{eq:j}
j_C^{\nu}=\bar{\Psi}\gamma^{\mu}\Psi=(\rho_R+\rho_L, \rho_R-\rho_L)
\end{equation}
is the current $2-$vector, and $\lambda^{(2)}_{\mu\nu}$ are interaction couplings. Here, we have introduced the subscript $C$ to represent the fact that these are the definition of current operators in the classical limit.

The action $S_{LL}$ has a $U_A(1)$ chiral symmetry with a parameter $\beta$ defined by 
\begin{equation}{\label{transf.eff}}
\left\{ \begin{array}{l}
\psi_R(x)\rightarrow \psi_R(x)'=e^{i\beta(x)}\psi_R(x) \\
\psi_L(x)\rightarrow \psi_L(x)'=e^{-i\beta(x)}\psi_L(x),
\end{array} \right.
\end{equation}
when the argument $\beta(x)$ is independent of the space and time coordinates $x$. In the case of a local transformation, where $\beta(x)$ varies 
non-trivially with 
coordinate $x$, the variation of the action $\delta S_{LL}= S_{LL,0}'-S_{LL,0}$ must be proportional to the derivative $\partial_\mu\beta$, thus written as 
\begin{equation}
\delta S_{LL}=-\int d^2x\, j_C^{5\mu}\partial_{\mu}\beta,
\end{equation}
which is used to define the chiral current as 
\begin{align}\label{eq:j5F}
j_C^{5\mu}=\bar{\Psi}\gamma^{\mu}\gamma_5\Psi
\end{align}
generated by the transformation where $\gamma_5=i\gamma^0\gamma^1$.
The variation of the action (following an integration by parts) considered above suggests the conservation of the chiral current (i.e. $\partial_{\mu}j_C^{5\mu}=0$) along the lines of Noether's theorem in classical mechanics.
The time-component of the current $j_C^{5,0}$ is a chiral charge density 
\begin{equation}\label{eq:rhoccb}
\rho_{c,cb}=\psi_R^\dagger\psi_R-\psi_L^\dagger\psi_L=\rho_R-\rho_L,
\end{equation}
whose form is particularly simple in this so-called chiral basis for fermions $\psi_{L,R}$.
However, as we will discuss below, a quantum-mechanical treatment of this action leads to a violation of the conservation law for the
chiral current, which is called the chiral anomaly.

The Tomonaga-Luttinger model $S_{LL}$ can be solved by representing the fermionic fields $\Psi$ in terms of a bosonic field $\Phi$, which is related to the current operator as 
\begin{align}\label{eq:jB}
&j^{\mu}=\frac{1}{\sqrt{\pi}}\epsilon^{\mu\nu}\partial_\nu \Phi, 
\end{align}
through a process known as bosonization~\cite{stone1994bosonization, Giamarchi2007}. However, the process of bosonization for such gapless models requires regularization to resolve the fate of the chiral current~\cite{Coleman1975}.
In the Lorentz-invariant case where $\lambda_{\mu\nu}^{(2)}=\lambda\eta_{\mu\nu}$, which is also referred to as the Thirring model~\cite{Georgi1971},
divergences in the quantum field theory can be regularized  
after transforming the action to Euclidean (i.e. Wick rotated) space~\cite{stone1994bosonization} by introducing an $2-$momentum cut-off. 
Applying this regularization to the Thirring model directly, one can check that (see Appendix \ref{apx:C} for details) the chiral current defined as the response to the chiral vector potential is related to the charge current as: 
\begin{align}\label{eq:j5}
&j^{5\mu}=-\epsilon^{\mu\nu}j_\nu=\frac{1}{\sqrt{\pi}}\partial^\mu \Phi,
\end{align} 
where $\epsilon^{\mu\nu}$ is the completely anti-symmetric unit tensor.
Note that we have not directly used the classical definitions of both charge and chiral currents (i.e. Eqs.~\ref{eq:j} and~\ref{eq:j5F}) but rather only the fact that the chiral current and charge current are related by a rotation. However, the chiral charge Eq.~\ref{eq:j5} is related to the classical chiral charge Eq.~\ref{eq:j5F} by a Euclidean point splitting regularization $j^{5\mu}(x)=\int d^2 x' \Gamma(|x'|)\bar{\Psi}(x)\gamma^{\mu}\gamma_5\Psi(x+x')$. The Lorentz invariance in Euclidean space is ensured by rotation invariance of $\Gamma$. Furthermore, $\Gamma(|x'|)\sim e^{-|x'|/\xi}$ decays to zero past a cut-off $\xi^{-1}$ and is normalized $\int d^2x'\Gamma(|x'|)=1$. In momentum space, this corresponds to a rotationally invariant vertex function $\Gamma(k)=(|k|^2+\xi^{-2})^{-1}$ and is similar to the heat-kernel regularization scheme that is used in the treatment of anomalies~\cite{Rylands2021}. Assuming the same regularization $\Gamma$ applies to $j^{\mu}$, as detailed in Appendix \ref{apx:C}, the relation between the chiral and regular currents (i.e. Eq.~\ref{eq:j5}) follows from the algebra of $\gamma$ matrices.
While the charge current $j^\mu$ in Eq.~\ref{eq:jB} is manifestly divergence-free, as detailed in the Appendix \ref{apx:C}, the divergence of the chiral current can be written in terms of the classical equation of motion for $\Phi$ as 
\begin{equation}{\label{Fujikawa_anomaly_1DEG}}
\left<\partial_\mu j^{5\mu}\right>=\frac{1}{\sqrt{\pi}}\partial_\mu\partial^{\mu}\Phi=\frac{1}{1+\lambda/\pi}\frac{e}{\pi}E.
\end{equation}
where $E=\partial_0A_1-\partial_1A_0$ is the electric field.
This is referred to the chiral anomaly equation in $1+1$D~\cite{Georgi1971}. Furthermore, 
since the r.h.s. depends on the interaction strength $\lambda$, the chiral anomaly is renormalized by interaction~\cite{Georgi1971}
in a way that is identical to that obtained directly from the Thirring model using either Pauli-Villars regularization or the Fujikawa method~\cite{Rylands2021}. 

Condensed matter systems without microscopic Lorentz-invariance allow a more general type of 
interaction $\lambda^{(2)}_{\mu\nu}=(g_4+g_2)\eta_{0\mu}\eta_{0\nu}+(g_4-g_2)\eta_{1\mu}\eta_{1\nu}$, while still preserving chiral symmetry.
In the absence of Lorentz-invariance, one can use a simple "point-splitting" regularization~\cite{stone1994bosonization} to regulate divergences in $S_{LL}$. The point-splitting approach involves replacing products of operators at the same space-time point $\rho_{R/L}(x,t)\rho_{R/L}(x,t)\rightarrow \rho_{R/L}(x-\epsilon,t)\rho_{R/L}(x+\epsilon,t)$ to slightly separated points in space at the same time. The correlators of the system can now be computed using a more 
conventional Hamiltonian operator-based approach~\cite{Coleman1975,Giamarchi2007},
which leads to a wave-equation for $\Phi$ 
\begin{equation}\label{TLeom}
u^{-1}\partial_t^2\Phi-u\partial_x^2\Phi=\frac{e K}{\sqrt{\pi}} E,
\end{equation}
where 
\begin{align}\label{uk}
&uK=1+\frac{g_4}{2\pi}-\frac{g_2}{2\pi},\,\quad u/K=1+\frac{g_4}{2\pi}+\frac{g_2}{2\pi}
\end{align}
(see Appendix \ref{apx:C} for a review of the derivation) ~\cite{Giamarchi2007}.
As reviewed in Appendix \ref{apx:C}, the bosonization identities~\cite{Giamarchi2007} can now be applied to the operator of the chiral charge density in Eq.~\ref{eq:rhoccb} to show that 
\begin{equation}\label{current-based-chiral}
\rho_{c,mb}(:=(\rho_R-\rho_L))=j/uK.
\end{equation}	
Note that this result for the chiral charge, though formally identical to Eq.~\ref{eq:rhoccb} actually includes a point-splitting regularization function $\Gamma$ similar to the relativistic result Eq.~\ref{eq:j5}. The difference is that in this case $\Gamma(x'_0,x'_1)\propto \delta(x'_0)$ i.e. $\Gamma$ is an equal time vertex function and only include point-splitting in space. The regulator $\Gamma$ is a momentum cut-off rather than the $2-$momentum 
 in the relativistic case. Unlike the relativistic case, the $\Gamma$ vertex cannot be applied to the conventional current, which must account for all the electrons in the system. However, since the occupation of fermions $\psi_k^\dagger\psi_k$ is fixed away from the Fermi wave-vector $|k-k_F|\gtrsim k_F$, as elaborated in Appendix~\ref{apx:C} the charge density $\rho=\rho_R+\rho_L$ can be written in terms of $\rho_{R,L}$. The second equality in Eq.~\ref{current-based-chiral} follows from charge conservation, as detailed in appendix~\ref{apx:C}. Note that the above relationship of $\rho_{c,mb}$ to the current $j$ is rather different compared to Eq.~\ref{eq:j5}, which is really dependent on a Lorentz-invariant regulator.
Applying this equation of chiral charge to the equation of motion leads to the following chiral anomaly equation:
\begin{equation}\label{bosonization_anomaly_1DEG}
\left<\partial_\mu j^{5\mu}_{mb}\right>=\frac{1}{\sqrt{\pi}}\partial_t\Pi+\partial_x j_{c,mb}=\frac{e}{\pi} E
\end{equation}
where $j_{c,mb}=u \rho/K$ is the chiral current, and $\Pi=\sqrt{\pi}(\rho_R-\rho_L)$ is the canonical conjugate field to $\Phi(x)$. Note that in contrast to the chiral anomaly equation (Eq.~\ref{Fujikawa_anomaly_1DEG}) resulting from a Lorentz-invariant regularization of the TL model, the above chiral anomaly equation has no interaction-based renormalization.
Incidentally, the chiral anomaly equations of the 1D edge for topological integer quantum Hall systems, based on either Eq.~\ref{Fujikawa_anomaly_1DEG} or Eq.~\ref{bosonization_anomaly_1DEG}, are both unrenormalized.
This is because in the latter formalism the two edges ($R$ and $L$) cannot scatter into each other in the IQHE, i.e., the $g_2$ term, $\rho_R\rho_L$, is missing; thus, the Luttinger parameter $K$ is always $1$ and the two chiral charges Eq.~\ref{current-based-chiral} and Eq.~\ref{eq:rhoccb} are equivalent. 

The above chiral anomaly equation (Eq.~\ref{bosonization_anomaly_1DEG}) is clearly different from relativistic chiral anomaly Eq.~\ref{Fujikawa_anomaly_1DEG} even in the formally Lorentz-invariant case $g_4=0$ (and $g_2=\lambda$).
Thus, the chiral anomaly in $(1+1)$D as well as the equation of motion for the current $j$, even in the presence of formal Lorentz-invariance, depend on regularization. 
Note that the difference between the chiral charges defined in Eqs. \ref{eq:j5} and \ref{current-based-chiral} originates from the regularization function $\Gamma$ despite the classical definition of the chiral charge (i.e. Eq.~\ref{eq:j5F}) being identical. In the classical field theory, the chiral charge is fixed by the Noether charge of the chiral symmetry. In the Lorentz-invariant Thirring model, response functions of all operators - including the chiral charge operator must be regularized in a manner that preserves Lorentz symmetry. In contrast, regularization schemes in condensed matter systems  typically uses a spatial point-splitting while retaining the equal time nature of observables. This is because most observables in condensed matter systems such as charge and current are products of equal-time operators. The application of these two different regularization schemes to the same classical current operator lead to Eqs. \ref{eq:rhoccb} and \ref{current-based-chiral}, respectively.

\subsection{Application to Interacting Spin-orbit Coupled Fermions\label{sec:2A}}
The simplest systems that one may hope to apply these ideas to would be an electron gas with parabolic dispersion and density-density interactions. The density-density interactions lead to the constraints $g_2=g_4$ and $uK=v_F=1$ 
in Eq.~\ref{uk} for the bosonized description of the system~\cite{Giamarchi2007}. Indeed, one can check that this produces the correct equation of motion for the current $j$ (i.e. Eq.~\ref{TLeom}) based on Galilean-invariance. Thus, the point-splitting 
regularization rather than the Lorentz-invariant one produces the correct conductivity. 
However, the chiral anomaly (i.e., Eq.~\ref{current-based-chiral}) is simply the current density.

The chiral charge $j^{5,0}$ can be distinct from the current in a Rashba spin-orbit coupled gas with density-density and spin-spin interactions.
A more general way to derive the coefficients $u,K$ is to directly analyze the equation of motion Eq.~\ref{TLeom}, which can be written as $(uK)^{-1}\partial_t j+(u/K)\partial_x\rho=(e/\pi) E$. In the limits $\omega\ll q$ and $\omega\gg q$, we can interpret $(u/K)$ as compressibility and $(uK)^{-1}$ as the inertia.
To compute these for the spin-orbit coupled model, consider the Hamiltonian 
\begin{align}
H_{soc}(k)&=(k-\alpha \sigma_z)^2-\alpha^2+\Gamma\sigma_x. 
\end{align}
For small $\Gamma$, the states near the Fermi energy have $k=\pm 2\alpha$ and are polarized along $\sigma_z=\pm 1$. In addition, let us consider a short-ranged interacting term 
\begin{align}
&H_{soc,int}=g_2\int dx \rho(x)^2\simeq g_2\int dx \rho_+(x)\rho_-(x).
\end{align} 
Note that in the small $\Gamma$ limit, $\rho_{\pm}\equiv\rho_{L,R}$ in the TL model. Therefore, we can apply the results Eq.~\ref{uk} with $g_4=0$ 
to this model and derive a non-trivial chiral charge $\rho_{c,mb}$, which is different from the current $j$, using Eq.~\ref{current-based-chiral} for this model.

\subsection{Relativistic Anomaly in Condensed Matter Systems\label{5}}
Since many of the low-energy degrees of freedom in condensed matter systems, such as the 1DEG, appear to have Lorentz-invariance based on the bosonized equation of motion Eq.~\ref{TLeom}, 
one may ask if the Lorentz-invariant anomaly (i.e. Eq.~\ref{Fujikawa_anomaly_1DEG}) applies to such systems in any sense. 
The results discussed earlier in this section show that different regularizations will generate different results of the chiral anomaly (i.e. Eqs.~\ref{Fujikawa_anomaly_1DEG},~\ref{bosonization_anomaly_1DEG}), some of which are renormalized by interactions and others are not. 
This issue can be resolved by refermionization~\cite{von1998bosonization}, where fermions can be constructed from a model of bosons.
Such a construction can be shown for the Sine-Gordon model~\cite{Coleman1975,mandelstam1975soliton}:
\begin{equation}\label{sg}
\begin{aligned}
S_{sg} = \int dt\, dx\, \frac{1}{2} \bigg[ \frac{1}{uK}(\partial_t \Phi)^2 
& - \frac{u}{K}(\partial_x \Phi)^2 \\
& \quad - u g \cos(2\sqrt{\pi}\Phi) \bigg]
\end{aligned}
\end{equation}
The Sine-Gordon model can be obtained as a low-energy description of a 1DEG using bosonization~\cite{Giamarchi2007} by adding a back-scattering term to the Luttinger-model $S_{LL}$, which is written as:
\begin{equation}
S_{LL,bs}=\int dt dx \, V_0 \cos(2k_F x) \psi^\dagger\psi.
\end{equation}
The Luttinger model $S_{LL}$ is discussed in Eq.~\ref{lut}. While the parameters $u,K$, and $g$ depend on the regularization scheme used (similar to the anomaly), the form of the model $S_{sg}$ is independent of regularization.
Interestingly, the term $S_{LL,bs}$ explicitly breaks the chiral symmetry (Eq.~\ref{transf.eff}) and the spectrum associated with $S_{sg}$ is gapped for $1<K<2$. The gap allows one to describe the excited states of the system in terms of solitons and antisolitons over a gapped vacuum~\cite{Coleman1975,mandelstam1975soliton}. The solitons and antisolitons are fermionic and are described by a massive Thirring model~\cite{Coleman1975}.
Furthermore, in the range of parameters $1<K<2$, the solitons and antisolitons are conserved excitations and can be used to define a soliton-based 
chiral charge:
\begin{equation}
\rho_{c,sb}=N_{s}+N_{\bar{s}}
\end{equation}
, where $N_{s(\bar{s})}$ is the number operator of (anti)solitons. 
The chiral symmetry in Eq.~\ref{transf.eff} is replaced by the integrability of $S_{sg}$ which leads to the conservation of the current density despite
the absence of an explicit chiral symmetry or momentum conservation. 

\begin{figure}[t]
\centering
\includegraphics[width=0.95\columnwidth]{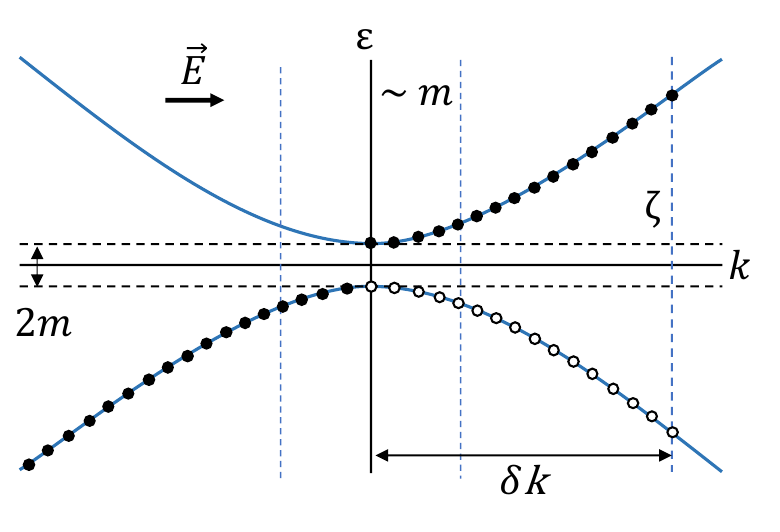}
\caption{Illustration of the chiral anomaly in the Sine-Gordon model. The energy spectrum is sketched as two curves with an energy gap of $2m$, where $m$ is the soliton mass. In the presence of an $E$, solitons and antisolitons are generated and are shown by black dots in the top band and white dots in the bottom band, respectively. The range between two dashed lines in the middle corresponds to low-speed moving (anti)solitons. The minimum highest-level momentum of soliton pairs is labeled by $\zeta$. }
\label{fig:sgEk}
\end{figure}

To understand the role of the explicit chiral-symmetry-breaking term, $V_0$, we consider the equation of motion 
\begin{equation}\label{eom}
\partial_tj+u^2\partial_x\rho+u^2Kg\sin(2\sqrt{\pi}\Phi)=uK\frac{eE}{\pi},
\end{equation}
where $j$ is given by Eq.~\ref{eq:jB} and the gap parameter $g\propto V_0$. This equation becomes equivalent to the gapless equation of motion Eq.~\ref{Fujikawa_anomaly_1DEG} at the limit $V_0\rightarrow 0$. 
Furthermore, considering the excitations of the Sine-Gordon model shown in Fig.~\ref{fig:sgEk}, where the gap is proportional to $m\sim g$,
we see that the limit $m\sim g\rightarrow 0$ implies that almost all solitons and antisolitons move at speed $u$. 
In this limit, the chiral charge is proportional to the current, i.e. $\rho_{c,sb}\simeq j/u$ (see detailed discussion in Appendix \ref{apx:B}). The chiral current is related to the chiral charge by the velocity of the modes, i.e., $j_{c,sb}\simeq u [N_{s}-N_{\bar{s}}]\simeq u\rho$. Applying this to Eq.~\ref{eom} (taking the limit $g\rightarrow 0$) leads to an anomaly equation 
\begin{equation}
\partial_t\rho_{c,sb}+\partial_x j_{c,sb}=K\frac{eE}{\pi}.
\end{equation}

This equation is essentially equivalent to the relativistic anomaly equation in Eq.~\ref{Fujikawa_anomaly_1DEG}. Note that the role of the mass is rather subtle, since we are using the limit $g\rightarrow 0$, which is the same limit considered in Eq.~\ref{Fujikawa_anomaly_1DEG}. However, the finite but vanishingly small $g$ is needed for the identification of $\rho_{c,sb}$ as a conserved chiral charge of solitons/antisolitons. This is because, as established earlier in the section, one cannot identify this charge with left-/right-moving microscopic fermions. The latter identification is dependent on regularization. From an experimental standpoint, the validity of the above equation assumes an electric field that is applied over a relatively short time (i.e. relatively high frequency) relative to the mass $m\sim g$. The integrability of the Sine-Gordon model, which leads to the conservation of the chiral charge $\rho_{c,sb}$ and also remarkably the corresponding physical charge current $j$, allows one to measure the chiral charge at a later point to check the validity of the anomaly. In practice, one could simply measure the rescaled current $j/u$ as the chiral charge. This does not require an actual backscattering potential. However, to check that this corresponds to a physical density, one would measure the soliton density at long times when the finite mass disperses the wave package so that individual solitons and antisolitons are separated. This check requires backscattering $V_0$ to create the dispersion of the wave packet and identify the individual charges. 

\section{Momentum-based chiral charge and unrenormalized anomaly in 1D\label{3}}
The regularization issues described above can be resolved by defining a chiral charge density in terms of momentum density. The basic idea is to treat the chiral symmetry as an emanant symmetry\cite{Cheng_2023} rather than an emergent symmetry, meaning that it emanates from the microscopic translation symmetry.
To understand this, 
let us first review how the mixed anomaly of translation symmetry and $U(1)$ symmetry is closely analogous to the chiral anomaly $U(1)$~\cite{Cho2017}.
In the low-energy regime, we can expand the fermionic field operator $\psi(x,t)$ in a neighborhood of $\pm k_F$ according to the relation:
\begin{equation}{\label{expan}}
\psi(x,t)\simeq e^{ik_Fx}\psi_R(x,t)+e^{-ik_Fx}\psi_L(x,t),
\end{equation}
where $k_F$ is the Fermi momentum and the right and left fermionic fields $\psi_R$ and $\psi_L$ that slowly vary in space and time.
With the help of this preliminary step, the effect of a space translation transformation can be written as
\begin{equation}
\left\{ \begin{array}{l}{\label{transf.mic}}
\psi_R(x,t)\rightarrow \psi_R(x,t)'=e^{ik_Fa(x,t)}\psi_R(x,t) \\
\psi_L(x,t)\rightarrow \psi_L(x,t)'=e^{-ik_Fa(x,t)}\psi_L(x,t).\\
\end{array} \right.
\end{equation}
Note that we ignore the shift of $\psi_{R/L}(x+a,t)$ by $a$ since these fields vary slowly compared to $k_F$. 
These transformations become equivalent to the chiral $U(1)$ transformation defined in Eq.(\ref{transf.eff}) for $\beta(x)=k_F a(x)$.

While the total momentum displays properties of a chiral anomaly in systems with discrete translation symmetry~\cite{Cho2017}, 
systems with continuous translation symmetry allow us to study a local momentum density field, which is analogous to a chiral charge density.
The chiral charge density is the Noether current associated with the chiral $U(1)$ symmetry in the systems described in Sec.~\ref{4}. Similarly, one can identify the low-energy limit of the canonical momentum density $T^{01}$, which is the Noether current associated with translation, as the analog chiral charge density for systems with finite bandwidth. 
However, unlike anomalous chiral $U(1)$ symmetry, translation symmetry is preserved under quantization.
Thus, the momentum current density $T^{\mu 1}$ is conserved without any anomaly. Therefore, the anomaly in this case is replaced by a mixed anomaly~\cite{Cho2017} that arises from the gauge dependence of $T^{\mu 1}$ under the $U(1)$ symmetry associated with charge conservation. In the case of continuous translation symmetry, this mixed anomaly can be transformed into a chiral anomaly equation for a locally gauge-invariant analog of the momentum density, as described below.

As a first example, we will start by computing the locally gauge-invariant momentum density, which we will call the kinetic momentum density, in the 1DEG model.
The action for the 1DEG model in the presence of an external gauge field in 1+1 is $S_{1DEG}=S_{1DEG,0}+S_{1DEG,int}$ with
\begin{equation}{\label{S20}}
S_{1DEG,0}=\int dx dt\, i\psi^\dagger D_0\psi-\frac{1}{2m}\bar{D}_1\psi^\dagger D_1\psi +\mu \psi^\dagger\psi,
\end{equation} 
where $D_\mu=\partial_\mu+ie A_\mu$ and $\mu$ is the chemical potential. For simplicity, we take the interacting part of the action as the density-density interaction with the form
\begin{equation}\label{S2int}
S_{1DEG,int}=\int dt dx dx' V(x-x') \rho(x)\rho(x'),
\end{equation}
where $\rho(x)=\psi^\dagger(x)\psi(x)$ is the fermionic number density with $V(x-x')$ being an interaction potential.
As detailed in Appendix \ref{apx:A}, the stress-energy tensor, which is the Noether current associated with translation, can be calculated considering the variation of the action under translation following Eq.~\ref{apx:eq:T}.
The spatial dependence of the vector potential $A_\mu(x,t)$ breaks translation-invariance and leads to a non-zero divergence of the stress-energy tensor
\begin{equation}\label{T_conti_A}
\partial_\mu T^{\mu\nu}=e\rho\partial^\nu A_0-ej\partial^\nu A_1
\end{equation}
, where $\rho$ and $j$ are the charge density and current, respectively.
The gauge dependence of $T^{\mu\nu}$ becomes apparent from the lack of gauge-invariance of the r.h.s. of the divergence 
of $T^{\mu\nu}$.
The gauge-invariance of the r.h.s. can be restored by rewriting 
Eq.~\ref{T_conti_A} as 
\begin{equation}\label{T_conti_F}
\partial_\mu(T^{\mu\nu}-ej^\mu A^\nu)=ej_\mu F^{\nu\mu}
\end{equation}
, where the notation $F^{\nu\mu}$ represents the electromagnetic field tensor. 
The above equation motivates the definition of a kinetic stress-energy tensor $K^{\mu\nu}$ as
\begin{equation}
K^{\mu\nu}=T^{\mu\nu}-ej^\mu A^\nu,
\end{equation}
which has a gauge-invariant divergence 
\begin{equation}\label{K_conti_F}
\partial_\mu K^{\mu\nu}=ej_\mu F^{\nu\mu}.
\end{equation}
Using the calculated $T^{\mu\nu}$ in Appendix \ref{apx:A}, the kinetic momentum density $K^{01}$ and the kinetic stress $K^{11}$ are given by 
the manifestly gauge-invariant form
\begin{align}
K^{01} &= -i \psi^\dagger D_1 \psi \label{K_01} \\
K^{11} &= i \psi^\dagger D_0 \psi 
+ \frac{1}{2m} \bar{D}_1 \psi^\dagger D_1 \psi \notag \\
&\quad\quad\quad + \left[ \int dx' V(x - x') \rho(x') + \mu \right] \rho(x). \label{eq:K01}
\end{align}

The electron gas model described by $S_{1DEG}$ is Galilean-invariant, which corresponds to the case $uK=1$ discussed in Sec.~\ref{sec:2A}, where the
chiral charge is exactly equal to the current. Similarly, the chiral charge resulting from $K^{01}$ which will be discussed in the following is also equal to the current. At the same time, the spin-orbit coupled system discussed in Sec.~\ref{sec:2A} requires higher derivative terms to describe. 
Therefore, we generalize the formalism to include higher spatial derivatives of the fermions with an action that is written as:
\begin{align}\label{sgnl}
S_{1DEG,h}=S_{1DEG}-\int dx dt \,V_{h}(\psi^\dagger,\psi,D_x\psi,D_x^2\psi,\cdots).
\end{align}
Such an action can describe
the spin-orbit coupled dispersion discussed in Sec.~\ref{sec:2A} when only the lower band fermions are considered.
The potential $V_h$ does not contain the time derivative term, since Hamilton is assumed to be time-independent. 
The continuity equation of $K^{\mu\nu}$ is still Eq.~\ref{K_conti_F}, but $K^{\mu\nu}$ is modified to
\begin{equation}
K^{\mu\nu}\rightarrow K^{\mu\nu}+C^{\mu\nu},
\end{equation}
where 
\begin{align}\label{C_munu}
C^{\mu\nu}&=\frac{\partial V_h}{\partial(\partial_\mu A^\rho)}\partial^\nu A^\rho
\nonumber\\
&+\left[\frac{\partial V_h}{\partial(\partial_\mu\partial_\rho A^\rho)}\partial^\nu\partial_\rho A^\rho
-\partial_\rho\left(\frac{\partial V_h}{\partial(\partial_\mu\partial_\rho A^\rho)}\right)\partial^\nu A^\rho\right]\notag\\
&+\cdots.
\end{align}
(See details in Appendix \ref{apx:A})

As shown in Appendix \ref{apx:A} the kinetic momentum $K^{01}$ written Eq.~\ref{eq:K01} receives no corrections from the higher derivative corrections $V_h$, although the expression for $K^{11}$ is now more complicated.
In addition, as detailed in Appendix \ref{apx:A}, the kinetic stress-energy tensor $K^{\mu\nu}$ (Eq.~\ref{K_gen}) is gauge-invariant when the Lagrangian density is gauge-invariant and therefore can be used to define a gauge-invariant chiral charge.

Since $K^{01}$ has dimensions of momentum, one can convert the kinetic stress-energy tensor to a chiral current using the relation 
\begin{equation}\label{momentum_based_chiral_charge}
j^{5\mu}_{mb}:=K^{\mu 1}/k_F.
\end{equation}
The corresponding momentum-based definition of the chiral charge density is
\begin{align}
\rho_{c,mb}:=K^{01}/k_F 
\end{align}
This resolution resolves the issue of identifying a nearly conserved chiral charge in a system without having to depend on regularization. 
The above result shows that in the systems of interest, the momentum density, 
which we know to be conserved in a system with continuous translation-invariance, can be served as a conserved chiral charge.
This is in contrast to the traditional chiral charge Eq.~\ref{eq:rhoccb}, which is only conserved in a TL model approximation, which requires a choice of regularization. Additionally, this definition is roughly consistent with Eq.~\ref{eq:rhoccb} since for a weakly interacting system, we expect the change in momentum $\delta T^{01}\sim k_F\rho_{c,cb}$.	
Finally, the conservation law for $j^{5,\mu}$ defined according to Eq.~\ref{momentum_based_chiral_charge}, which can be obtained from Eq.~\ref{K_conti_F}, is an anomaly equation that is identical to Eq.~\ref{bosonization_anomaly_1DEG} and is written as
\begin{equation}\label{momentum-anomaly}
\left<\partial_\mu j^{5\mu}_{mb}\right>=e\frac{n}{k_F}F^{10}=\frac{e}{\pi}E,
\end{equation}
where we have simplified the r.h.s. of Eq.~\ref{K_conti_F} $\left<ej_\mu F^{1\mu}\right>=e n E$.
For the last equality, we have used the Luttinger relation~\cite{Luttinger1960,Blagoev1997} $n=k_F/\pi$ is the average number density, and $E=F^{10}$ is the external electric field, which is the only non-zero component of the electromagnetic tensor in $1+1$D. This result is consistent with the non-Lorentz-invariant bosonization result in Eq.~\ref{bosonization_anomaly_1DEG}. 
In contrast to the Lorentz-invariant renormalized anomaly in Eq.~\ref{Fujikawa_anomaly_1DEG},
this anomaly is not renormalized by interaction. Note that the anomalous non-vanishing of the r.h.s of Eq.~\ref{momentum-anomaly} in this case arises from the fact that $K^{1\mu}$ differs from $T^{1\mu}$, the conserved canonical stress-energy tensor, which is not gauge-invariant. This is thus the result of a mixed anomaly~\cite{Cho2017} rather than the conventional chiral anomaly arising from regularization.

\subsection{Application: Chiral Charge Conservation between Strong-Coupling Luttinger Liquids with Different \texorpdfstring{$K$}{K}}
The momentum-based chiral anomaly has a compelling and practical application in the context of a junction between two Luttinger liquids (LLs) with different Luttinger parameters \( K \). The Hamiltonian can be expressed as
\begin{equation}
H_{K}=\int dx\frac{u(x)}{2}\left[K(x)\Pi^2+K(x)^{-1}(\partial_x\Phi)^2\right], 
\end{equation}
where the boson field \(\Phi\) is defined in terms of the charge density \(\rho=\partial_x \Phi/\sqrt{\pi}\) as in Eq.\ref{eq:rho}. \(K(x)\) describes a junction that smoothly varies from \(K_-\) to \(K_+\). The rest of the bosonization process including the resulting current operator \( j \) remains the same \( \left( j = -\partial_t \Phi/\sqrt{\pi} = u(x)K(x)\Pi/\sqrt{\pi} \right) \) and its conservation law remains the same as in Appendix.\ref{apx:C}. 
Using canonical commutation relations, as before, leads to a slightly more general equation of motion for the field \(\Phi\):
\begin{equation}
\frac{\partial_t^2\Phi}{u(x)K(x)}-\partial_x\left(\frac{u(x)}{K(x)}\partial_x\Phi\right)=0. 
\end{equation}
Integrating over \(x\) on both sides, the chiral charge conservation law around this junction is obtained:
\begin{equation}{\label{eqn:positiondpdtchiralcharge}}
\partial_t Q_c=\partial_t\left(\int dx\, j/u(x)K(x)\right)=0,
\end{equation}
where \(Q_c=\int dx \,\Pi/\sqrt{\pi}\) represents the momentum-based chiral charge (i.e. Eq.~\ref{current-based-chiral}), as opposed to the Lorentz-based definition in Eq.~\ref{eq:rhoccb}.

As a notable application, consider a recent experiment~\cite{Cohen2023} that observed a quantized \(e^2/2h\) conductance from a \(\nu_-=1/3\) fractional quantum Hall edge (\(K_-=1/3\)) tunneling into a $\nu_+=1$ integer quantum Hall edge (\(K_+=1\)) in the strong coupling limit. This quantum point contact can be modeled as a quantum wire with a spatially varying Luttinger parameter \(K(x)\) just as discussed above\cite{thomas2024quantumhalltransformerquantum}. The scattering pattern at the junction can be determined by using normal and chiral charge conservation. Suppose that the incoming charge packet from the FQH edge carries \(q_{\nu_-,in}\) units of the basic charge \(e\) and moves to the right, the reflected charge packet carries \(q_{\nu_-,rf}\) units and moves to the left, and the transmitted particle carries \(q_{\nu_+,out}\)(integer) units and moves to the IQH system. Based on conventional charge conservation, we obtain
\begin{equation}
q_{\nu_-,in} = q_{\nu_-,rf} + q_{\nu_+,out}.
\end{equation}
Next, consider the chiral charge, $Q_c$, of these three charge packets. Using the chiral charge definition in Eq.~\ref{eqn:positiondpdtchiralcharge}, we find 
the corresponding chiral charges of each of the packets to be: \(Q_{c,\nu_-,in}=K_-^{-1}eq_{\nu_-,in}\), \(Q_{c,\nu_-,rf}=-K_-^{-1}eq_{\nu_-,rf}\), and \(Q_{c,\nu_+,out}=K_+^{-1}eq_{\nu_+,out}\), respectively. We note that to obtain these results we needed the corresponding current operators for each of the packets, i.e. $\tilde{j}_{\nu_-,in}(0)=u_-eq_{\nu_-,in}$, $\tilde{j}_{\nu_-,rf}(0)=-u_-eq_{\nu_-,rf}$, and $\tilde{j}_{\nu_+,out}(0)=u_+eq_{\nu_+,out}$, respectively. The conservation of chiral charge then leads to the equation:
\begin{equation}
K_-^{-1}q_{\nu_-,in} = -K_-^{-1}q_{\nu_-,rf} + K_+^{-1}q_{\nu_+,out}
\end{equation}
Solving the equations for the conservation of charge and chiral charge yields the following:
\begin{align}
q_{\nu_-,in} &= \frac{K_+^{-1}+K_-^{-1}}{2K_-^{-1}} q_{\nu_+,out}\\
q_{\nu_-,rf} &= \frac{K_+^{-1}-K_-^{-1}}{2K_-^{-1}} q_{\nu_+,out}.
\end{align}
This indicates that the scattering patterns are restricted to this specific configuration. Consequently, the corresponding dc conductance $G$ of this Andreev reflection-like process is\cite{Sandler1998}
\begin{equation}
G=\frac{e^2}{h}T_{K_-,K_+}=\frac{e^2}{h}K_-\frac{q_{\nu_+,out}}{q_{\nu_-,in}}=\frac{2e^2}{h}(K_-^{-1}+K_+^{-1})^{-1},
\end{equation}
where $T_{K_-,K_+}$ is the transmission coefficient.

Referring back to the experimental settings, with the given $K$, we obtain $q_{1/3,in} = (2/3)\, q_{1,out}$, $q_{1/3,rf} = (-1/3)\, q_{\nu_+,out}$, and the dc conductance $G$ is $e^2/2h$. This result is consistent with the measured conductance in experiment ~\cite{Cohen2023} and provides an explanation of the predicted quantum Hall transformer conductance~\cite{chklovskii1998consequences} using a chiral charge-conserving Andreev reflection-like process.

This chiral charge conservation law (i.e., Eq.~\ref{eqn:positiondpdtchiralcharge}) directly results from the canonical momentum conservation of $\Pi$. Although we did not initially define the chiral charge in this subsection using microscopic translation symmetry, the bosonization process depends on microscopic details through regularization. Consequently, the effective conservation of canonical momentum in bosonization still originates from the microscopic effective translation symmetry resulting from a smooth potential in the UV description of the QPC, rendering the chiral charge conservation to be an emanant symmetry~\cite{Cheng_2023}. 
Thus, the chiral charge in this case remains "momentum"-based corresponding to the microscopic translation symmetry.
Adding backscattering terms that break the effective translation symmetry in the UV could invalidate the chiral charge conservation law.

\section{Stability of Chiral anomaly in 3D\label{6}}
The chiral anomaly for non-interacting 3D systems can be understood~\cite{Nielsen1983,Armitage2018} by focusing on the LL spectra in a magnetic field and applying the 1D results already discussed. However, this decoupling into 1D systems is not preserved in the presence of interactions. Additionally, as discussed in previous sections, the definition of a chiral charge in 1D is ambiguous unless a momentum-based approach to chiral charge is used. On the other hand, 3D Weyl systems away from the Weyl point can be viewed as a topological Fermi liquid~\cite{Haldane2004} where the low energy response is described by free quasiparticles. The chiral charge associated with each Weyl point can be defined in terms of the Luttinger volume of each Fermi pocket without any reference to the regularization required in the 1D case. In Sec.~\ref{sec:6a}, we generalize the momentum-based anomaly equation discussed in Sec.~\ref{3} to the 3D case. In Sec.~\ref{sec:6b}, we show that this anomaly equation is equivalent to the anomaly equation based on the Luttinger-volume-based chiral charge. We contrast this in Sec.~\ref{sec:6c} with the direct computation of the change in Luttinger volume using chiral kinetic theory~\cite{Son2013}. In Sec.~\ref{sec:6d}, we describe how the momentum-based approach continues to apply in the low-temperature Landau level limit beyond the application of chiral kinetic theory. Finally, in Sec.~\ref{sec:6e} we propose a pump-probe measurement of the Luttinger-volume-based chiral anomaly.

\subsection{Momentum anomaly in \texorpdfstring{$3+1$D}{3+1D}\label{sec:6a}}
For this section, we will assume that the separation between the Weyl points is small compared to the lattice constant, so that the dispersion can be approximated by a continuum model~\cite{Armitage2018} with conserved momentum. As will become clear in Sec.~\ref{sec:6d}, 
while the lattice momentum in the $z-$direction is important to define the total momentum $P_z$ in the system, it does not play a role in the chiral anomaly. Additionally, the role of this lattice momentum can be eliminated (regularized) by defining the momentum relative to a
trivial insulator that can be defined for the same model.
The momentum-based chiral charge in 1+1D, i.e., Eq.~\ref{momentum_based_chiral_charge}, can be generalized to 3+1D in a straightforward way as:
\begin{equation}{\label{momentum_rho3D}}
j^{\mu,3D}_{c,mb}:=K^{\mu z}_0/k_0
\end{equation}
, where $2k_0$ is the separation between the Weyl nodes and $K^{\mu z}_0$ is the kinetic stress-energy tensor associated with the electrons below the Weyl points.
We choose the axes $z$ to be along the direction connecting the two Weyl points. 
For the purposes of the validity of Fermi-liquid theory, we will assume that the Fermi level is away from the Weyl point. This means that $K^{\mu z}_0$ generically differs from $K^{\mu z}$, which is the kinetic stress tensor of the topological Fermi liquid~\cite{Haldane2004}.

Following the kinetic momentum density approach to the chiral anomaly discussed in Sec.~\ref{3} for the 1D case, 
we start with the momentum conservation equation for a charged fluid (as in the $1+1$D case) in an electric field assuming the magnetic field along the $z$-direction, i.e. 
\begin{equation}\label{eq:pz_continuity_3D}
\partial_0 \langle K^{0z}\rangle +\partial_i \langle K^{iz}\rangle =e\langle \rho\rangle E_z.
\end{equation}
Note that the charge density $\rho$, unlike the 1D case, is not related to $k_0$.
In fact, the relevant part of the charge density $\rho$, here, is the contribution that is linear in $B$ due to the Streda relation:
\begin{equation}\label{eq:sigmaH}
\lim_{B\to 0}\left.\frac{\partial \rho}{\partial B}\right|_{\mu}=\frac{e^2}{4\pi^2}Q_z^H
\end{equation}
, where $\sigma^{ab}_H=(e^2/4\pi^2) \epsilon^{abc} Q^H_c$ is the intrinsic Hall conductivity. As was shown by Haldane~\cite{Haldane2004}, the vector $Q_c^H$ can be written as an integral over the Berry curvature over the Fermi surface
\begin{align}\label{eq:Haldane}
	Q^H_c=\frac{1}{2\pi}\int ds^\mu \wedge ds^\nu \mathcal{F}_{\mu\nu}(\boldsymbol{k}(\boldsymbol{s}))k_c(\boldsymbol{s}),
\end{align}
where $\wedge$ is the wedge product, representing the oriented area, \( \mathcal{F}_{\mu\nu}(\boldsymbol{k}(\boldsymbol{s})) \) is the Berry curvature 2-form on the Fermi surface and $\boldsymbol{s}$ parameterizes the Fermi surface.
The anomalous Hall conductance is thus a well-defined property of a strongly interacting Fermi liquid.

The above relation also implies that for Weyl points with a net Berry flux, the vector $Q_z^H$ approaches $Q^H_z\rightarrow 2k_0$ in the limit of the Fermi energy approaching the Weyl point. When the Fermi energy is away from the Weyl point, one can apply the Streda relation in Eq.~\ref{eq:sigmaH} to the states between the Weyl point and the Fermi surface to calculate the magnetic-field-induced correction to the charge density as
\begin{align}\label{eq:rhoW}
&\delta \rho_W=\frac{e^2}{4\pi^2}\delta Q_z^HB=\frac{e^2}{4\pi^2}(Q^H_z-2k_0)B.
\end{align}
Here, $\delta \rho_W$ is the $B$-dependent correction to the extra density of electrons associated with the Fermi level being away from the Weyl point. 
This allows us to identify the electric field-induced change in momentum of the electrons between the Fermi surface and the Weyl point $K^{\mu z}_0$:
\begin{equation}\label{dK0}
\partial_\mu (\langle K^{\mu z}-K^{\mu z}_0\rangle)=E_z\delta \rho_W=\frac{e^2}{4\pi^2}E_zB\delta Q^H_z.
\end{equation}
Subtracting this change from the total change in momentum (i.e., Eq.~\ref{eq:pz_continuity_3D}) and applying Eq.~\ref{eq:sigmaH}, we get 
\begin{equation}\label{eq:K0}
\partial_0 \langle K^{0z}_0\rangle+\partial_i\langle K^{iz}_0\rangle=\frac{e^2}{2\pi^2}k_0 E_z B.
\end{equation}
Combining this observation with the definition of chiral charge in Eq.~\ref{momentum_rho3D} leads to the chiral anomaly equation 
\begin{equation}\label{eq:anomaly_3D}
\partial_0 \langle \rho^{3D}_{c,mb}\rangle+\partial_i \langle j^{i,3D}_{c,mb}\rangle=\frac{e^2}{2\pi^2}(\boldsymbol{E}\cdot \boldsymbol{B})
\end{equation}
We note that we have not explicitly written an expression for the chiral current $K^{iz}_0$ since it is not necessary to define the chiral anomaly coefficient. However, this is expected to be a direct generalization of Eq.~\ref{eq:K0}.
We also note that this momentum-based chiral anomaly is unrenormalized as expected, although the momentum $K^{0z}_0$ and $k_0$ individually might be renormalized by interactions.

The inclusion of Coulomb interactions between the electrons, in the absence of other trivial Fermi surfaces, constrains the total electron density of the bulk 3D system to be constant instead of the constant chemical potential assumed so far. This results in a correction of the chemical potential in response to the magnetic field $B$ because of the Streda relation 
\begin{equation}
\delta \mu_B= -\frac{e^2}{4\pi^2}Q_z^H B/N(0),
\end{equation}
since the spectral flow-induced change in charge density at constant chemical potential (i.e., Eq.~\ref{eq:sigmaH}) is compensated by $\delta \mu_B$.
This magnetic field-induced change in chemical potential is by itself an interesting variant of the chiral anomaly since this is another way in which the
electromagnetic field can affect the total charge of the Weyl point. As a result of this change, Eq.~\ref{eq:rhoW} for the charge density around the Weyl point is modified to 
\begin{equation}\label{eq:rhoWC}
	\delta \rho_W^{(Coul.)}=\frac{e^2}{4\pi^2}\delta Q_z^H B+N(0)\delta \mu_B =-\frac{e^2}{2\pi^2}k_0B
\end{equation}
Interestingly, the charge neutrality constraint implies that $\rho$ in Eq.~\ref{eq:pz_continuity_3D} must vanish. 
Despite this, the application of Eq.~\ref{dK0} leads to the conclusion that the chiral anomaly equation Eq.~\ref{eq:anomaly_3D} is preserved even when the charge-neutral limit is enforced by Coulomb interactions. This is one rather stringent test of the robustness of the anomaly to interactions. 

\subsection{Luttinger Volume-based Chiral Charge\label{sec:6b}}
The electric field $E$ typically changes the momentum of the quasiparticles on the Fermi surface by shifting the Fermi surface, resulting in a momentum contribution $E_z\delta\rho_W$. But, as seen in Eq.~\ref{dK0}, the change in the Weyl part of the momentum $K^{0z}_0$ explicitly excludes this piece.
Therefore, the change $\delta K^{0z}_0$ must arise from an electric-field-induced transfer of charge between the two Weyl points
that are separated by the wave vector $2k_0$ so that $k_0^{-1}\delta K^{0z}_0=\delta\rho^{3D}_{c,mb}=\delta \rho_{w}$ is the number of electrons transferred between the Weyl points. 
Such a transfer of charge between the Weyl points has also been derived using chiral kinetic theory ~\cite{Stephanov2012,Son2013,Haldane2004}
and is explicitly written as:
\begin{equation}\label{eq:LV_chiral_anomaly}
\frac{\partial \rho_{w}}{\partial t} + \nabla\cdot \boldsymbol{j}^{3D}_{c}= \frac{e^2}{2\pi^2} (\boldsymbol{E}\cdot \boldsymbol{B})
\end{equation}
in the low-temperature limit, in which scatterings between two Weyl nodes can be suppressed and neglected.
In fact, this non-conservation of the Weyl point charge $\rho_{w}$ is the original definition of the chiral anomaly in $3+1$D~\cite{Adler1969}, 
which however requires regularization.
However, in the finite electron density limit discussed here, this chiral charge at the Weyl points $\rho_w$, which is equivalent to the momentum-based chiral charge $\rho^{3D}_{c,mb}$, can be defined as a Fermi-surface property through the Luttinger volume as: 
\begin{equation}{\label{rhoFS}}
\rho^{3D}_{c,FS}:=\sum_{\alpha=\pm}\gamma_\alpha\int_{\overline{FS}_\alpha}\frac{d^3k}{(2\pi)^3}
\end{equation}
, where $FS_\alpha$ represents the separate Fermi surfaces around two Weyl nodes, and $\gamma_\alpha=\pm$ indicates the chirality of the Weyl nodes. Eq. \ref{rhoFS} represents a form of chiral charge, that is well-defined even for interacting systems where
Fermi surfaces are defined as $G^{-1}(k,\omega=0,\mu)=0$, where $G$ is the single-particle Green's function. The Luttinger volume provides an interaction robust definition of a chiral charge that does not rely on any form of regularization that is not available in $1+1$D or in the strong magnetic field limit with only the chiral Landau level being occupied where the chiral anomaly is often discussed~\cite{Nielsen1983}. 

\begin{figure}[t]
\centering
\includegraphics[width=0.95\columnwidth]{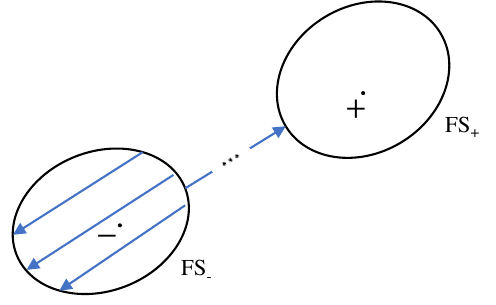}
\caption{Schematic of a scattering pattern around the Fermi surface at finite $T$.}
\label{fig.scattering}
\end{figure}

It should be noted that the equivalence of the momentum-based chiral anomaly $\rho^{3D}_{c,mb}$ and the chiral charge $\rho^{3D}_{c,FS}$ breaks down when the Fermi surface is at finite temperature and sufficiently far from the Weyl point.
This is because at finite temperature $T$, the scattering of quasiparticles around the Fermi surface which conserve momentum but not chiral charge as shown in Fig.\ref{fig.scattering} will contribute to relaxation of the chiral charge $\rho^{3D}_{c,FS}$ at rate $\sim T^2$. 
However, such scatterings have no influence on the total momentum; the momentum-based chiral anomaly provides immunity against scattering. 
The rates of such relaxation will be highly suppressed at any finite temperature since this will have to be a very correlated multiparticle process. 
On the other hand, momentum can be affected even by intra-Weyl node disorder scattering, which does not affect $\rho^{3D}_{c,FS}$.
In the case of discrete translation symmetry, Umklapp scatterings from the lattice will eliminate the momentum-based chiral anomaly. The rates of such relaxation processes are expected to be suppressed in the low-density limit, where the Weyl points are close to each other relative to the Brillouin zone.

\subsection{Chiral kinetic theory\label{sec:6c}}
The results of the arguments in Secs.~\ref{sec:6a} and ~\ref{sec:6b} can be checked explicitly using the chiral kinetic theory of the quasiparticles~\cite{son2012berry} 
of the topological Fermi liquid~\cite{Haldane2004}. This is valid in the limit that the magnetic field is small enough so that the associated cyclotron frequency is much lower than the thermal energy, i.e., $\omega_B\ll k_BT $.	
In the Fermi liquid limit, this chiral kinetic theory has been successfully used to understand the properties of topological Fermi liquids~\cite{Stephanov2012, Duval2006, Son2013, Xiao2005}. In the following we will use it to validate the properties of the momentum-based chiral charge as well as the measured chiral charge discussed in Secs.~\ref{sec:6a} and ~\ref{sec:6b}.

The central equation of kinetic theory is the Boltzmann equation:
\begin{equation}
\frac{\partial f}{\partial t}+\frac{\partial f}{\partial \boldsymbol{r}}\cdot \dot{ \boldsymbol{r}}+\frac{\partial f}{\partial \boldsymbol{k}}\cdot \dot{ \boldsymbol{k}}=I_{coll}=-\frac{\delta f}{\tau}\label{eq:f}
\end{equation}
, where $f=f( \boldsymbol{r}, \boldsymbol{k},t)$ is the quasiparticle distribution function.
The quasiparticles can only change their state and be equilibrated through collisions, which are approximated by the relaxation time approximation with a constant collision frequency $1/\tau$.

The equations of motion for quasiparticles in the presence of an anomalous velocity~\cite{karplus1954hall} are written as:
\begin{eqnarray}
\dot{ \boldsymbol{k}}&=&e \boldsymbol{E}+e\dot{ \boldsymbol{r}}\times \boldsymbol{B}\\
\dot{ \boldsymbol{r}}&=&\boldsymbol{v_k}+\dot{ \boldsymbol{k}}\times \boldsymbol{\Omega}_{ \boldsymbol{k}}\label{eq:dr}
\end{eqnarray} 
, where $A_{ \boldsymbol{k}}=i\left<u_{ \boldsymbol{k}}|\nabla u_{ \boldsymbol{k}}\right>$, $ \boldsymbol{\Omega}_{ \boldsymbol{k}}=\nabla \times A_{ \boldsymbol{k}}$ is the Berry curvature, and the group velocity $\boldsymbol{v_k}=\nabla_k \varepsilon$. 
Due to the Berry curvature term in the above equations leads to 
a violation of Liouville's theorem by the r.h.s. of Eq.~\ref{eq:f}
~\cite{Xiao2005} This can be rectified by introducing a phase space density factor $G_{ \boldsymbol{k}}=(1+e \boldsymbol{B}\cdot \boldsymbol{\Omega}_{ \boldsymbol{k}})^2$, which satisfies an equation of motion\cite{Stephanov2012}
\begin{equation}
\frac{\partial}{\partial t}\sqrt{G_k}+\frac{\partial}{\partial \boldsymbol{r}}\cdot (\sqrt{G_k}\dot{ \boldsymbol{r}})+\frac{\partial}{\partial \boldsymbol{k}}\cdot (\sqrt{G_k}\dot{ \boldsymbol{k}})=e^2( \boldsymbol{E}\cdot \boldsymbol{B})\nabla\cdot \boldsymbol{\Omega}_{ \boldsymbol{k}}.
\end{equation}
We can define a conserved phase-space density $\tilde{f}=\sqrt{G_k}f$ that obeys Liouville's theorem by combining the phase-space density $\sqrt{G_k}$ with the distribution function $f$, which now obeys the conservation law
\begin{equation}
\;\;\frac{\partial}{\partial t}\tilde{f}+\frac{\partial}{\partial \boldsymbol{r}}\cdot (\tilde{f}\dot{ \boldsymbol{r}})+\frac{\partial}{\partial \boldsymbol{k}}\cdot (\tilde{f}\dot{ \boldsymbol{k}})=e^2( \boldsymbol{E}\cdot \boldsymbol{B})\nabla\cdot \boldsymbol{\Omega}_{ \boldsymbol{k}}f +\tilde{I}_{coll}\;
\end{equation}
, where $\tilde{I}_{coll}=\sqrt{G_k}I_{coll}$. 
Using the above equation of motion for the phase space density $\tilde{f}$, we can find the continuity equation for the Luttinger volume-based chiral charge in Sec.\ref{sec:6b} and the kinetic momentum 
\begin{align}\label{eq:P}
\boldsymbol{P}=\int _{\boldsymbol k} \boldsymbol{k} \tilde{f}
\end{align}
in Sec.\ref{sec:6a}. The result of the first one (Eq.\ref{eq:LV_chiral_anomaly}) has been already discussed in many references and Sec.\ref{sec:6b}. The second one, with respect to the kinetic momentum, is written as 
\begin{equation}{\label{boltz_p}}
\frac{\partial}{\partial t} \boldsymbol{P}+\nabla\cdot \vec{\boldsymbol{K}}=\frac{e^2}{4\pi^2}( \boldsymbol{E}\cdot \boldsymbol{B})\boldsymbol{Q}^H + \int_{\boldsymbol{k}} \left(e \boldsymbol{E} + e \boldsymbol{v}_{\boldsymbol{k}} \times \boldsymbol{B} \right) f.
\end{equation}
, where 
\begin{equation}
\vec{\boldsymbol{K}}=\int_{\boldsymbol k} \left(\boldsymbol{v_k}+e \boldsymbol{E}\times \boldsymbol{\Omega}_{ \boldsymbol{k}}+e\left(\boldsymbol{v_k}\cdot \boldsymbol{\Omega}_{ \boldsymbol{k}}\right) \boldsymbol{B}\right) \boldsymbol{k}f
\end{equation} 
, a dyadic tensor, is the stress tensor, and 
\begin{equation}\label{eq:Q_h}
\boldsymbol{Q}^H =4\pi^2\int_{\boldsymbol k} \boldsymbol{\Omega}_{ \boldsymbol{k}} f=\sum_\alpha \boldsymbol{Q}^H_\alpha= \frac{1}{2\pi}\sum_\alpha\int_{S_\alpha} d^2 \mathcal{F}\,\boldsymbol{k}
\end{equation}
, which is just the vector associated with the AQH conductivity in the Streda formula Eq.\ref{eq:sigmaH}.
 The collision term will not contribute if we assume that collisions are elastic. Incidentally, this equation is general for any condensed matter system, and interactions will not be affected. 

To obtain the momentum-based chiral anomaly equation from Sec.\ref{sec:6a}, we still consider the $z$-direction component of the kinetic stress-energy tensor and assume the magnetic field also along $z$. Eq.\ref{boltz_p} will be simplified:
\begin{equation}\label{bolt_p_proj}
\frac{\partial}{\partial t} P^z+\nabla\cdot \vec{K}^z=\frac{e^2}{2\pi^2}( \boldsymbol{E}\cdot \boldsymbol{B})Q^H_z.
\end{equation}
Restricting the integrals in Eqs.~\ref{eq:P} and ~\ref{eq:Q_h} above the Weyl point leads to a variation of the above equation:
\begin{equation}\label{dp0}
\quad\;\frac{\partial}{\partial t} (P^z-P^z_0)+\nabla\cdot (\vec{K}^z-\vec{K}^z_0)=\frac{e^2}{2\pi^2}( \boldsymbol{E}\cdot \boldsymbol{B})\delta Q^H_z,\quad
\end{equation}
where $P^z$ and $Q^H_z$ are replaced by $(P^z-P^z_0)$ and $\delta Q^H_z=Q_z^H-2k_0$, which are the momentum and anomalous Hall contributions of electrons above the Weyl point, respectively. Not that $P^z_0$ and $2k_0$ are intrinsic momenta and the anomalous Hall contributions of the Weyl point. 
This provides a more quantitative understanding of Eq.~\ref{dK0} from Sec.~\ref{sec:6a}. Combining Eq.~\ref{bolt_p_proj} with Eq.~\ref{dp0} leads us back to the chiral anomaly equation:
\begin{equation}{\label{eq:momentum_based_continuity_3D_high_T}}
\frac{\partial}{\partial t} P_0^z/k_0+\nabla\cdot \vec{K}_0^z/k_0=\frac{e^2}{2\pi^2}( \boldsymbol{E}\cdot \boldsymbol{B}),
\end{equation}
Thus, the analysis of the contribution of the Weyl point to the momenta $P^z_0$ provides a more microscopic derivation (i.e. in terms of FL quasiparticles) 
of the anomaly equation Eq.~\ref{eq:anomaly_3D}. 

As an side, the chiral kinetic theory can also compute the chiral magnetic effect(CME). The current along the $B$-direction can be written as\cite{Stephanov2012}
\begin{align}\label{eq:jcme}
\boldsymbol{j}_{\rm{CME}}=e^2\boldsymbol{B}\int_{\boldsymbol k} f \left(\boldsymbol{v_k}\cdot \boldsymbol{\Omega}_{ \boldsymbol{k}}\right)=\frac{e^2\boldsymbol{B}}{4\pi^2}\Delta \mu,
\end{align}
where
\begin{equation}
\Delta \mu=\frac{1}{2\pi}\sum_\alpha \int_{S_\alpha} d^2 \mathcal{F}\,\varepsilon(\boldsymbol{k})
\end{equation}
(See details in Appendix \ref{apx:D}) is the deviation of the chemical potential from an equilibrium state. Note that since states far away from the Weyl point can contribute to the current, it is important to consider the chiral magnetic current $j_{CME}$ relative to an equilibrium state with vanishing current~\cite{vazifeh2013electromagnetic}. If we assume that the electron distribution functions have equilibrium forms in the individual valleys, $\Delta \mu$ is exactly the chemical potential difference at the Fermi level, and Eq.\ref{eq:jcme} is consistent with references\cite{Son2013, Armitage2018}. Eq.\ref{eq:jcme} is robust since it depends only on Fermi surfaces. Hence, the conductivity along the magnetic field
direction, $\sigma_{zz}$, is modified as\cite{Son2013,Armitage2018}
\begin{equation}
\sigma_{zz}(B)=\sigma_{zz}(0)+\frac{e^4B^2\tau_a}{4\pi^4g(\epsilon_F)}
\end{equation}
, where $g(\epsilon_F)$ is the density of state and $\tau_a$ is the inter-node scattering time.
The above conductance depends on the parameters $\tau_a$ and $g(\epsilon_F)$. On the other hand, the chiral magnetic current $\boldsymbol{j}_{CME}$ in terms of $\Delta\mu$ as given by Eq.~\ref{eq:jcme} is universal for a topological Fermi liquid and not renormalized by interactions. While the above equation for dc conductivity depends on inter-valley scattering at rate $\tau_a^{-1}$, one could consider the response to an ac electric field $E(\omega)$ at frequency $\omega\gg \tau_a^{-1}$. In this case, the ac current response should satisfy Eq.~\ref{eq:jcme} and lead to a corresponding fluctuation of $\Delta\mu$ at frequency $\omega$ provided $\omega$ is slower than the intra-valley equilibration rate. The chemical potential difference $\Delta\mu$ can, in principle, be measured by the dissipative optical response at frequency $\omega_1\sim \Delta\mu+v_s k_0$, where $v_s k_0$ is the minimum phonon frequency of intervalley scattering from acoustic phonons with sound velocity $v_s$ and wave vector $k_0$, which is the separation of Weyl points. The phonon frequency is a bit difficult to estimate. However, one can consider the change in $\Delta\mu$ with varying electric field amplitudes to verify the chiral magnetic effect.

\subsection{Low Temperature Landau Level Limit\label{sec:6d}}
\subsubsection{Model}
The discussion so far has assumed a temperature that is high relative to the small applied magnetic field. However, all the conclusions are carried out equally well at nearly zero temperature. 
The low temperature for a non-interacting Weyl system in a magnetic field is described by the chiral Landau level~\cite{Nielsen1983}.
To illustrate this, we will compute the non-interacting Landau levels for the continuum Weyl toy model
~\cite{Armitage2018}, which is written as:
\begin{equation}\label{eq:Weyl_model}
\begin{aligned}
H_{wl}&=H_0+H_{R}\\
&=(k^2-k_0^2)\sigma_z+v(k_y \sigma_x -k_x \sigma_y)
\end{aligned}
\end{equation}
, where $k^2=k_x^2+k_y^2+k_z^2$ and $v$ is the Rashba coupling constant. The energy spectrum is indeed the Weyl type in the low-energy regime, which has two touching Weyl points at $\boldsymbol{k}=(0,0,\pm k_0)$ when $v\neq 0$.

\begin{figure}
\centering
\includegraphics[width=0.95\columnwidth]{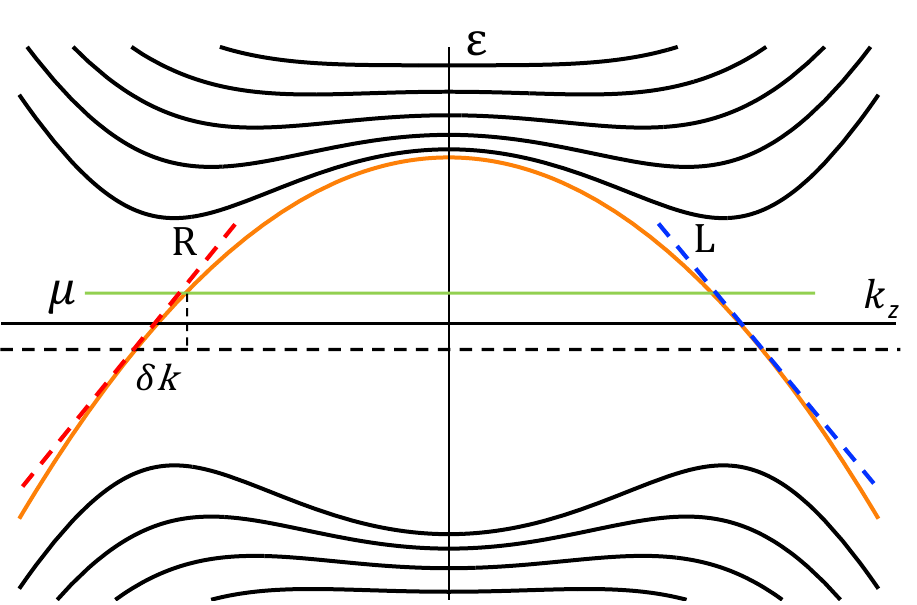}
\caption{Band structure of Weyl model ($k_0=1$) in the Sec.\ref{sec:6d} with a small spin-obit coupling $v=1.5$ under the magnetic field $B=0.1$. The chemical potentials are set away from the Weyl nodes, and the low energy states are basically multiply LLs.}
\label{fig:weyl}
\end{figure}

To study the chiral anomaly, when applying the magnetic field $B$ 
along the $z$-direction, we calculate the LL spectrum of the Hamiltonian $H_{wl}$ in the presence of a magnetic field, as shown in Fig.~\ref{fig:weyl}. While the magnetic field preserves the angular momentum, the spin-orbit coupling mixes the $(\sigma_z=1,n)$ and $(\sigma_z=-1,n+1)$ (where $n$ labels the conventional Landau levels without spin-orbit) and opens a gap in the spectrum. However, one of the Landau levels $(\sigma_z=-1,n=0)$ does not have a partner to hybridize and remains gapless. This Landau level, which is referred to as the chiral Landau level, is responsible for the chiral anomaly in an applied electric field within the non-interacting limit~\cite{Nielsen1983}. Note that the gap in the non-chiral LLs in Fig.~\ref{fig:weyl} actually occurs at negative energies as opposed to the previous calculation of LL spectra of individual Weyl cones~\cite{Nielsen1983, Armitage2018, Burkov2018}.
This difference is a manifestation of the Streda formula i.e. Eq.~\ref{eq:sigmaH}, which, as discussed in Sec.~\ref{sec:6a} is critical for understanding the chiral anomaly from a momentum conservation perspective.

\subsubsection{Momentum-based Chiral Charge\label{CLLmom}}
The chiral anomaly is conventionally defined, using the argument of Nielsen and Ninomiya~\cite{Nielsen1983}, as non-conservation of chiral charge can be determined from the LL spectrum in Fig.~\ref{fig:weyl}
for the non-interacting case where the chemical potential is in the gap of the non-chiral LLs. This argument also leads to a current that appears to be renormalized by interactions~\cite{Rylands2021}, which, similar to the $1+1$D results discussed in Sec.~\ref{4}, depends on the regularization scheme used for the calculation.

However, the subtleties of defining a chiral charge discussed in Sec.~\ref{4} arise when only the Fermi level intersects just the chiral LL, and therefore the other LLs are important in defining chiral charge. 
Therefore, it is beneficial to consider the chiral anomaly away from the Weyl points, where several non-chiral LLs are also occupied. These non-chiral 
LLs do not contribute to the change in chiral charge in response to an electric field and therefore do not modify the chiral anomaly~\cite{Armitage2018}.
These non-chiral LLs do, however, affect the momentum-based definition of chiral charge because they contribute to momentum change under an electric field.
This turns out to be not a problem because, similar to the case discussed in Sec.~\ref{sec:6a}, the contribution of non-chiral LLs as well as the electrons in the chiral LL above the Weyl point to the change in momentum can be calculated directly from the density of electrons in the respective LLs. For small magnetic fields corresponding to $\omega_B\ll E_F$,
this density is given by the FS volume.
Subtracting the momentum contribution of electrons above the gap of the non-chiral LLs leaves only the momentum contribution associated with a transfer of electrons from one Weyl point to the other. 
This can be estimated to be $2k_0 (B/2\pi)(E_z\delta t)$ where $E_z\delta t$ is the change in momentum of the electrons and $B/2\pi$ is the degeneracy per unit area of the chiral Landau level. The factor $2k_0$ arises from the length of the unoccupied segment (or occupied by holes) of the chiral Landau level in Fig.~\ref{fig:weyl}. Since $2k_0$ is the distance between the Fermi points on the chiral LL when the chiral LL crosses the gap of the non-chiral LL, the momentum density can be interpreted as a chiral charge that is consistent with the result Eq.~\ref{eq:anomaly_3D}. The chiral LL in Fig.~\ref{fig:weyl} leaves an ambiguity in the value of $2k_0$ to be in the range where the chiral LL crosses the gap. This ambiguity drops out in the limit of small $B$ since the gap also vanishes in this limit.

Fully filled LLs (including a fully filled chiral Landau level) will not contribute to the continuity equation Eq.~\ref{eq:pz_continuity_3D}. 
This is because the total crystalline momentum along $k_z$ generated by a slowly time-varying external electric field $E_z$ that increases a vector potential $\delta A_z=E_z\delta T=\Phi_0/L_z$, changes the momentum on the r.h.s of Eq.\ref{eq:pz_continuity_3D} by a primitive reciprocal lattice vector. Note that the lattice does play an essential role in regularizing the otherwise divergent momentum. However, 
one would obtain the same result from other regularizations such as by considering the difference in momentum between this chiral system and a variant of Rq.~\ref{eq:Weyl_model} with a trivial gap generated by a $\sigma_x$ term instead of the spin-orbit term.

\subsubsection{Interaction effects}
The effect of interactions can be included in a Fermi liquid theory in this framework if one goes away from the Weyl point (relative to the cyclotron frequency) i.e. $E_F\gg \omega_B$. In this limit and in the presence of a small magnetic field $\boldsymbol{B} = B\hat{\boldsymbol n}$, the quasiparticles near the FS follow semiclassical orbits (given by Eq.~\ref{eq:dr} with $\boldsymbol{E}=0$) which are confined to planes with constant parallel momentum $\boldsymbol{k}\cdot\hat{\boldsymbol{n}}=k$, with a long scattering lifetime. 
The Fermi-surface orbit in $k$-space on the FS labelled by $\alpha$, has a cross-sectional area $A_\alpha(k)$ 
that is set by a quantization condition~\cite{ashcroft2022solid}
\begin{equation}\label{eq:LLquant}
A_\alpha(k^n_{F\alpha})l_B^2-\phi_\alpha(k^n_{F\alpha})=2\pi (n+\frac12),
\end{equation}
where $l_B^2=1/eB$, and $n$ is an integer that labels the Landau orbits (i.e. bands in Fig.~\ref{fig:weyl}) and $\phi_\alpha (k)$ is the Berry phase of the quasiparticle on the phase-space orbit at $\boldsymbol{k}\cdot\hat{\boldsymbol{n}}=k$ on $FS_\alpha$.
The Landau band in Fig.~\ref{fig:weyl} with an extremum closest to the Fermi level represents the pair of Landau orbits on the FS that is closest to the maximum area orbit at $k=k_{max}$ of the FS. These are the orbits that contribute to quantum oscillations and are used to determine the size of the Fermi surface~\cite{ashcroft2022solid}. The solution with maximum $|k|$ is the chiral Landau level, which is a result of the difference in Berry phase $2\pi c_\alpha=\phi_\alpha(k_1)-\phi_\alpha(k_2)$,
where $k_1>k_{max}>k_2$ represents the range of $k$ of $FS_\alpha$ and $c_\alpha=\pm 1$ is the Chern number of Fermi surface around the Weyl point $\alpha$. 
The resulting LL spectrum near the FS is essentially identical to the non-interacting one seen in Fig.~\ref{fig:weyl}, except that the energy spectrum away from the Fermi energy is ill-defined because of the imaginary part of the self-energy.
The response to an external electric field $\boldsymbol{E}$ in the time period $\delta t$ is determined by the states near the Fermi energy, 
which is therefore identical to the non-interacting case discussed in the previous paragraph and is described by Eq.~\ref{eq:anomaly_3D}. 

\subsection{Measuring the Luttinger-volume-based chiral anomaly\label{sec:6e}}
The quantum oscillations that arise from the extremal Landau bands at $k=k_{max}$ discussed in the previous subsection can, in principle, be used
to detect the Luttinger-volume-based chiral anomaly discussed in Sec.~\ref{sec:6b} by measuring the chiral charge transfer in a pump-probe measurement.
The central idea is to use quantum oscillations to measure the anomaly-induced change in the Fermi wave vector as is done in magneto-transport (i.e. the Shubnikov-de Hass effect)~\cite{ashcroft2022solid}. This effect relies on the fact that low-frequency longitudinal conductivity would show dips associated with peaks in the density of states (DOS) when the Fermi level crosses a band extremum in Fig.~\ref{fig:weyl}. This corresponds to an extremal LL being allowed on the FS. For simplicity we will assume that the Fermi surface is nearly spherical near the Weyl point, so that the electron number density in the Weyl point $\alpha$, $N_\alpha=A_\alpha(k_{max})^{3/2}/6\pi^{7/2}$, can be written in terms of the extremal area $A_\alpha(k_{max})$.

Our protocol for measuring the Luttinger volume-based chiral anomaly discussed in Sec.~\ref{sec:6b} consists of three steps.

Step (1): Choose a magnetic field $B$ such that the FS quasiparticle DOS is at a local maximum, as evident from a minimum in the measured longitudinal conductance. In our model, this corresponds to $k_{max}$ satisfying Eq.~\ref{eq:LLquant} for an integer $n$.

Step (2): Apply a "pump" electric field $E_z$ for a short time $\delta t$ and wait for a relaxation time $\tau$.

Step (3): Measure the longitudinal conductance using a "probe" field to see if the conductance is near the minimum value.
Repeat step (2) with a different $\delta t$ utill one reaches
near the minimum value. The result of the measurement is $\Lambda=eE_z\delta T$, which transfers the system from one conductance minimum to another.

For a small magnetic field $B$, the change $\delta k$ in $k_{max}$ is insignificant, so that we can ignore the Berry phase term and simplify Eq.~\ref{eq:LLquant} to $\delta A_\alpha=2\pi \alpha eB$. The corresponding change in number density is $\delta N_\alpha=4 \alpha A_\alpha(k_{max})\delta k/(2\pi)^3=eB\alpha \sqrt{A_\alpha(k_{max})}/2\pi^{5/2}$.
Assuming an unrenormalized chiral anomaly and using Eq.\ref{eq:LV_chiral_anomaly}, the predicted $\Lambda$ would be 
\begin{align}
&\Lambda_{predict}=eE_z\delta T=\frac{2\pi^2}{e}\frac{\delta (N_+-N_-)}{B}=2\sqrt{\frac{A_\alpha(k_{max})}{\pi}}.
\end{align}
The measured value of $\Lambda$ determines the Luttinger-volume-based chiral charge transfer between the two Fermi surfaces, which based on the results in this section should not be renormalized by interactions or other complexities of the system.

\section{Summary and Discussion}
In summary, we start by illustrating the difficulties of applying chiral anomaly calculations from relativistic field theories by comparing the results to those from Hamiltonian bosonization in Sec.~\ref{4}.
We show that these ambiguities in the chiral anomaly equation are resolved by replacing the chiral anomaly by a mixed anomaly between continuous translation and $U(1)$ charge symmetry~\cite{Cho2017}.

The chiral anomaly equation here is qualitatively consistent with the corresponding effective action term derived previously using the correspondence between the t'Hooft and mixed anomalies~\cite{wang2021emergent,hughes2024gapless}.
We find that the mixed anomaly approach in $1+1$D leads to a chiral charge which is defined as the kinetic momentum divided by the Fermi wave vector $k_0$. The electromagnetic force term in the kinetic stress-energy continuity equation corresponds to the anomalous i.e. chiral symmetry breaking term.
As a compelling application, we examine the Andreev reflection-like process at the junction between a \(\nu = 1\) IQH edge and a \(\nu = 1/3\) FQH edge. Using conventional and chiral charge conservation, we can determine the unique scattering pattern, which is consistent with experimental observations.

The resulting chiral anomaly equation is unrenormalized by interactions. We find that it is still possible to realize an interaction-renormalized Lorentz-invariant anomaly in Luttinger liquids with weak backscattering. Such systems can be approximated by Sine-Gordon models. The chiral charge in this case must be chosen to be a fermionic representation based on solitons and antisolitons.

We then apply the mixed anomaly technique~\cite{Cho2017} to the $(3+1)$D Weyl systems. Similarly to previous work mapping the t'Hooft anomaly to an effective action term in a mixed anomaly representation~\cite{wang2021emergent}, we can use the Streda formula to show that the anomaly based on the difference of charge density around Weyl pockets is robust and unrenormalized. This is quite different from the chiral magnetic effect~\cite{Armitage2018},
whose manifestations appear with various non-universal prefactors that are renormalized by interactions. While some of these prefactors such as density of states or Fermi velocity, may be determined from other measurements, it is unclear whether the chiral magnetic effect is the only contributor to these currents. As an example of such subtleties, we note that the homogeneous finite frequency Hall conductivity $\sigma_{xy}(\omega)$ of a Weyl material~\cite{Rylands2021} was found, using heat kernel regularization, to contain a second-order term ($O(\omega^2)$) that was entirely determined by interaction strength. On the other hand, a numerical calculation of the ac Hall conductivity in a non-interacting model of a Weyl material shows a non-zero and non-universal second-order term even in the absence of interactions (see Appendix \ref{apx:E}). Similarly, a direct interpretation of the chiral magnetic effect in terms of the axion term in effective action~\cite{chen2013axion,goswami2015axial} suggests a magnetic field-induced current at a finite magnetic field. While such response terms can occur for finite frequency magnetic fields with open boundary conditions~\cite{alavirad2016role} as well as in the form of the gyrotropic magnetic effect~\cite{ma2015chiral,zhong2016gyrotropic}, this response is not observable in the presence of any physical equilibriation~\cite{vazifeh2013electromagnetic}. On the other hand, as discussed at the end of Sec.~\ref{sec:6c}, the unrenormalized CME relation (i.e. Eq.~\ref{eq:jcme}) can in principle be measured by a pump-probe experiment. However, this chiral magnetic effect is separate~\cite{Armitage2018} from the anomaly, which is the violation of charge conservation in an individual Weyl pocket. This observable in $(3+1)D$, which does not survive the ambiguities in $(1+1)D$, can be defined in terms of the Luttinger volume of the topological Fermi surface~\cite{Haldane2004}. While this is substantially more difficult to measure than a current response, we propose a rather experimentally intricate pump-probe measurement to directly measure the changes in the Luttinger volume in response to an electromagnetic field.

It should be noted that, in contrast to prior work~\cite{Cho2017,wang2021emergent}, to derive the chiral anomaly equations of motion we consider a mixed anomaly between $U(1)$ charge and continuous space translation. This leads to us having to use a regularization for the kinetic momentum discussed in Sec.~\ref{CLLmom} where we subtract off the contribution of the closest gapped insulator. This only makes sense if the Weyl points are close in momentum space relative to the total Brillouin zone, as in the model described in Sec.~\ref{sec:6d}. This is quite different from the Chern insulator stack model of a Weyl semimetal~\cite{Armitage2018} and the approach to the anomaly considered here may not apply to that case. On the other hand, the chiral kinetic theory~\cite{son2012berry} for the charge density of each pocket would still imply a quantized change in chiral charge, 
thus predicting a robust chiral anomaly. However, this relies on the full Fermi liquid emergent IR symmetry of the conservation of quasiparticles~\cite{else2021Non-Fermi}. This would be broken at finite temperature scattering of quasiparticles and Umklapp scattering from the edge of the Brillouin zone can eliminate the conservation of chiral charge. The continuum case considered here is more robust because the anomaly fundamentally is defined in terms of a continuum momentum that is conserved in a "low density" limit. Of course, disorder scattering ~\cite{Armitage2018} complicates this entire discussion, and there cannot be a precisely defined notion of chiral anomaly in this case. The realistic definition of the chiral anomaly necessarily requires considering the appropriate assumptions about the scaling with disorder~\cite{Son2013,burkov2015chiral, Armitage2018}.

The definition of the anomaly in this work has been in terms of equations of motion similar to the original works~\cite{adler1969axial,bell1969nuovo, Georgi1971} 
rather than the effective action approach for the t'Hooft anomaly of the emergent IR chiral symmetry~\cite{else2021Non-Fermi}. In fact, the topological chiral anomaly terms in the effective action of a microscopic mixed anomaly of $U(1)$ gauge and translation have already been worked out~\cite{wang2021emergent}. This approach is more general compared to the results presented here because it allows a general translation of known t'Hooft anomalies of various low-energy theories to mixed-anomaly-based responses in an effective action. In the context of Weyl, this work~\cite{wang2021emergent} clearly shows signatures
of the chiral Landau level and, presumably, other features of the chiral anomaly. In contrast, the present work only presents the equation of motion responses of the chiral anomaly, which are sometimes subtle to extract from an effective action as has been discussed in the previous paragraph. A direct approach to derive the universal aspects of the microscopic response function from the effective action~\cite{wang2021emergent} would be an interesting future direction that would provide a general anomaly formalism applicable to condensed matter systems. 

In this work, we have focused on the case of a single pair of Weyl points, where the mixed-anomaly formulation is the most transparent. The situation becomes more subtle in systems with multiple Weyl points. The momentum-based definition of chiral charge relies on connecting the anomaly to momentum conservation, which has only three components. Therefore, it does not naturally extend to separately conserved chiral charges associated with many Weyl nodes. Indeed, time-reversal-symmetric Weyl materials are automatically excluded, as they exhibit no intrinsic Hall response $\delta Q^H$, which is essential for the momentum-based anomaly. There are special cases, such as systems with four Weyl points and broken time-reversal symmetry, where one can choose the $z$-direction perpendicular to the line connecting a pair of opposite chirality nodes, so that this pair does not contribute to the momentum-based anomaly. The remaining pair then allows a limited extension of our approach. Beyond such cases, fully generalizing the momentum-based framework to arbitrary multi-node Weyl systems remains nontrivial and is left for future work.

In lattice systems with discrete translational symmetry, the situation differs from the continuum case. Adding a lattice to a Luttinger liquid, for example, introduces higher-order scattering processes—such as Umklapp terms—that formally break the emergent continuous translation symmetry. However, these processes are typically irrelevant at low energies (e.g., at low temperatures or away from commensurate fillings), so an effective continuous translation symmetry can still emerge, and the framework developed in this work remains applicable. In this discrete setting, the continuous momentum $k$ should be replaced by the crystal momentum $k_{\text{Bloch}}$, defined modulo reciprocal lattice vectors $G$. As a result, the chiral charge is conserved only up to reciprocal lattice translations, and the anomaly-free condition appears as a quantized pumping of crystal momentum across the Brillouin zone. A key subtlety here is that the total momentum transfer $K_0$ in Eq.~\ref{momentum_rho3D} must remain smaller than $G$ in order for the definition of chiral charge to be valid; this condition can be satisfied for sufficiently small (vanishing in the thermodynamic limit) applied electric and magnetic fields $E$ and $B$. This viewpoint naturally connects to the mixed anomaly between $U(1)$ charge and non-on-site discrete symmetry discussed in Ref.~\cite{Cheng_2023}. That said, in the presence of a lattice the chiral anomaly equation cannot be written in the standard continuum form, and may instead correspond to a discrete version of the anomaly. Exploring this possibility could be an interesting direction for future work.

When considering interaction effects, it is worth noting that our analysis is explicitly carried out within the Fermi liquid framework or using momentum conservation, where interactions are already included at the quasiparticle level. This allows us to avoid dealing with diagrammatic corrections directly. While some previous works have addressed similar questions using diagrammatic approaches—such as Ref.~\cite{giuliani2021anomaly}, which employed Ward identities to show that, in certain lattice models, the anomaly response is insensitive to the interaction strength and form—those results do not directly apply to our continuum Weyl model. Nevertheless, for our continuum setup, it is likely that some of these interaction-induced contributions (e.g., diagrammatic corrections) would cancel as well, and exploring this possibility could be another interesting direction for future work.

\begin{acknowledgments}
The authors acknowledge useful discussions with Colin Rylands and Victor Galitski, which motivate this work. We also thank Maissam Barkeshli for the discussions that told us about the relationship between an anomaly and a non-onsite symmetry. We thank Anton Burkov for alerting us to the mixed anomaly approach in Weyl semimetals~\cite{wang2021emergent} and Joel Moore for valuable comments in the later stages of this work. We acknowledge support from the Joint Quantum Institute co-directors fund and the Laboratory for Physical Sciences through the Condensed Matter Theory Center.
\end{acknowledgments}

\appendix
\section{Details of Bosonization\label{apx:C}}
\subsection{Lorentz-invariance Constrained Chiral Anomaly}
Let us now review the chiral anomaly in the relativistic case where we have chosen $\lambda_{\mu\nu}^{(2)}=\lambda\eta_{\mu\nu}$ so that $S_{LL}$ 
is then the Lorentz-invariant Thirring model~\cite{Georgi1971}. In this case, the collective mode velocity $u$ continues to match the Fermi velocity $u=v_F=1$, and current and density have 
the same units. 	 The chiral anomaly equation for this so-called Thirring model\cite{Georgi1971},
which was derived using perturbation theory together with a Lorentz-invariant regulator, shows a renormalization.
Here we use bosonization to derive the anomaly equation in a way where Lorentz and non-Lorentz-invariant results can directly be compared.

We consider the Euclidean (i.e. Wick rotated) space-time~\cite{stone1994bosonization} so that the point-splitting expansion is manifestly rotation (i.e. Lorentz)-invariant in the Wick rotated $(1+1)$D plane. Using this scheme of normal ordering the chiral fermionic 
operators $\psi_{R, L}(x,t)$ can be written as vertex operators of chiral bosonic operators $\Phi_{R, L}(x,t)$, which in turn can be used to 
define the bosonic field $\Phi(x,t)$~\cite{stone1994bosonization}. 
Applying the standard bosonization identities in Euclidean space~\cite{stone1994bosonization}
for replacing the Fermions in the kinetic term, $S_{1,0}$, we obtain 
\begin{equation}\label{eq:BosThirring}
S_{1,0}=\frac{1}{2K}\int (\partial_\mu\Phi)^2
\end{equation}
, where $x_0=t$ and $K=1$.
Similar use of the bosonization identities~\cite{stone1994bosonization} leads to related expressions for the chiral current 
and the $U(1)$ charge current, which can be written as 
\begin{align}\label{eq:j5App}
&j^{\mu}=\frac{1}{\sqrt{\pi}}\epsilon^{\mu\nu}\partial_\nu \Phi \,\quad j^{5\mu}=\frac{1}{\sqrt{\pi}}\partial^\mu \Phi,
\end{align}
where $\epsilon^{\mu\nu}$ is the completely anti-symmetric unit tensor. The role of regularization in this relation will elaborated in the next subsection. 

Applying these identities to $S_{1,int}$ leads to an additional contribution to Eq.~\ref{eq:BosThirring}, so that the $K$ factor
depends on the interaction strength in the fermionic model as 
\begin{equation}
K^{-1}=1+\lambda/\pi,
\end{equation} 
where $\lambda=g_2$ and $g_4=0$ because of Lorentz-invariance.

The coupling to an external vector potential $A_\mu$ is included through a term 
\begin{equation}\label{eq:lem}
\mathcal{L}_{em}=-e j^\mu A_\mu=-\frac{e}{\sqrt{\pi}}A_\mu\epsilon^{\mu\nu}\partial_\nu\Phi.
\end{equation}
While the charge current $j^\mu$ in Eq.~\ref{eq:j5App}	is manifestly conserved, the divergence of the chiral current can be written in terms 
of the classical equation of motion for $\Phi$ as 
\begin{equation}{\label{Fujikawa_anomaly_1DEGApp}}
\left<\partial_\mu j^{5\mu}\right>=\partial_\mu\partial^{\mu}\Phi/\sqrt{\pi}=\frac{1}{1+\lambda/\pi}\frac{e}{\pi}E.
\end{equation}
where $E=\partial_0A_1-\partial_1A_0$ is the electric field.
This result is obtained by using the expression for the chiral current Eq.~\ref{eq:j5App} and then combining Eq.~\ref{eq:BosThirring} with Eq.~\ref{eq:lem}.
This shows that the chiral charge, in contrast to the classical result is not conserved since the right-hand side is non-zero and is proportional to the electric field $E$. This is referred to as the chiral anomaly equation. Furthermore, 
since the right hand side depends on the interaction strength $\lambda$, the chiral anomaly is renormalized by interaction\cite{Georgi1971}. 
This result is identical to that obtained directly from the Thirring model using either Pauli-Villars regularization or the Fujikawa method~\cite{Rylands2021}. 

\subsection{Anomaly with Non-Lorentz-invariant Point-splitting Regularization\label{subsec4.b}}
Bosonization of the Luttinger model $S_{LL}$ (from Eq.~\ref{lut} and ~\ref{lutint}) can also be approached from a Hamiltonian perspective that is more appropriate for condensed matter systems that break Lorentz-invariance~\cite{stone1994bosonization}. Historically this was developed in parallel with the 
Euclidean formulation in the last subsection. This formalism is simpler because it directly uses operators in a Hamiltonian formalism. The point-splitting in space-time is now replaced by point-splitting in real space. This allows using the definition of the chiral charge density Eq.~\ref{eq:rhoccb} as well as the corresponding equation for total density as operator equations. In fact, the chiral charge density operators $\rho_{R, L}$ in $S_{LL}$ are promoted to operators, which, with the appropriate point-splitting obey the algebra~\cite{stone1994bosonization}
\begin{align}\label{apx:eq:comm}
&[\rho_a(x),\rho_b(x')]=-\frac{i}{2\pi}a\delta_{ab}\partial_x\delta(x-x').
\end{align}
Using the above commutation relation, the TL Hamiltonian can be written entirely in terms of chiral density operators
\begin{align}
H &=\int dx \, i\bar{\Psi}(\gamma^1\partial_x-i \gamma^0 \varphi
)\Psi+2g_2\rho_R\rho_L+g_4(\rho^2_R+\rho^2_L)\nonumber\\
&=\int dx \, (1+g_4)(\rho^2_R+\rho^2_L)+2g_2 \rho_R\rho_L+\varphi \rho,
\end{align}
where $\varphi$ is the electric potential. 

We can bosonize the above model using the operator version of Eq.~\ref{eq:j} for the density operator written as
\begin{align}\label{eq:rho}
\rho=\rho_R+\rho_L=\frac{1}{\sqrt{\pi}}\partial_x \Phi, 
\end{align}
where $\Phi$ is the bosonic field. Defining $\Pi=(\rho_R-\rho_L)\sqrt{\pi}$ as the canonically conjugate field to $\Phi(x)$, which is based on the commutation relation Eq.\ref{apx:eq:comm}, 
the above Hamiltonian can be written in bosonized form~\cite{Giamarchi2007}
\begin{equation}
H_{1DEG}=\int dx\frac{1}{2}\left(uK\Pi^2+\frac{u}{K}(\partial_x\Phi)^2\right)+E \Phi
\end{equation}
, where 
\begin{align}\label{ukApp}
&uK=1+\frac{g_4}{2\pi}-\frac{g_2}{2\pi}\\
&u/K=1+\frac{g_4}{2\pi}+\frac{g_2}{2\pi}
\end{align}
and $E=-\partial_x\varphi$ is the electric field. 
The current operator $j$ is now defined to be 
\begin{equation}\label{eq:jH}
j=-\frac{1}{\sqrt{\pi}}\partial_t \Phi=\frac{1}{\sqrt{\pi}}uK\Pi
\end{equation}
so that it satisfies the continuity equation for the charge. 
Note that the two equations Eq.~\ref{eq:rho} and Eq.~\ref{eq:jH} are direct operator analogs of Eq.~\ref{eq:j} 
except that $j$ is no longer related to $\rho_{R,L}$ in the same way as an operator.
This relation can be used to define the chiral charge in terms of the current operator 
\begin{equation}\label{current-based-chiralApp}
\rho_{c,mb}=(\rho_R-\rho_L)=\frac{1}{\sqrt{\pi}}\Pi=j/uK=-\frac{1}{\sqrt{\pi}}\partial_t \Phi/uK.
\end{equation}
Applying this definition to the equation of motion for $\Phi$ (Eq.~\ref{Fujikawa_anomaly_1DEG}) we obtain the chiral 
anomaly equation 
\begin{equation}\label{bosonization_anomaly_1DEGApp}
\partial_t \rho_{c,mb}+\partial_x j_{c,mb}=\frac{1}{\sqrt{\pi}}\partial_t\Pi+\partial_x j_{c,mb}=\frac{e}{\pi} E
\end{equation}
where $j_{c,mb}=u \rho/K$ is the chiral current. Note that in contrast to the 
chiral anomaly equation (Eq.~\ref{Fujikawa_anomaly_1DEGApp}) resulting from a Lorentz-invariant regularization of the TL model, the above chiral anomaly equation has no interaction-based renormalization. 

Generally, the regularized chiral anomaly can be derived by considering a regularized version of the chiral symmetry transformation in Eq.~\ref{transf.eff} where \begin{equation}
\beta(x)\rightarrow\beta(x,x')\equiv \int d^2x'\beta(x)\Gamma(x-x')T_{x'-x},
\end{equation}
and $T_{x'-x}\psi_{R(L)}(x)=\psi_{R(L)}(x')$ is the translation operator. This generalied chiral transformation does not affect the unitarity if $\Gamma$ is real since the translation operator $T_{x'-x}$ is Hermitian. 
The non-locality of $\Gamma$ allows one to introduce cutoffs such as momentum cutoffs (which break Lorentz) or heat kernel regulators in Euclidean space that are Lorentz-invariant. Accordingly, we will have the regularized chiral current:
\begin{equation}
j^{5\mu}(x)=\int d^2 x' \Gamma(x')\bar{\Psi}(x)\gamma^{\mu}\gamma_5\Psi(x+x').    
\end{equation}
In this form, $\Gamma$ appears like a vertex function for the current operator.

In the system with Lorentz symmetry (i.e. relativistic systems) in imaginary time, then the kernel depends only on the distance between the two points, and therefore
$\Gamma(x') = \Gamma(|x'|)$.
In the classical case $\Gamma(x)=\delta(x)$, but is a localized function in the regularized quantum case where the symmetry has a weaker effect on the higher momentum modes. As a result $\Gamma(k)\sim M^2/(k^2+M^2)$ regularizes the divergence of the loop integrals and leads to a finite answer provided $\Gamma$ is applied to also the current operator $j^{\mu}$ as well as the interaction $g$. This regularization is similar in many ways to Pauli-Villars regularization with a mass $M$ scalar field. Since the same regularization enters $j^{\mu}$ as $j^{5\mu}$, the $\gamma$ matrix algebra can be used to check $j^{5\mu}=\epsilon_{\mu\nu}j^{\nu}$ consistent with Eq.~\ref{eq:j5App}. In the system without Lorentz symmetry (e.g. condensed matter systems), the kernel $\Gamma(x_0',x_1')$ is $\propto \delta(x_0')$ instead of $\delta(x)$. In momentum space, $\Gamma(k_0,k_1)\sim M^2/(k_1^2+M^2)$. This is essentially equivalent to a momentum cutoff so that the $\mu=0$ component of the above equation is valid as the operator equation Eq.~\ref{current-based-chiralApp}.  Thus, because of the different $\Gamma$, the regularized chiral charge and chiral anomaly is different as expected. The choice of these different vertex functions lead to distinct chiral currents Eqs.~\ref{eq:j5} and ~\ref{current-based-chiral} respectively.

As a side note, we note that in the Galilean-invariant case, the Hamiltonian $H_{1DEG}$ can be written in terms of the current 
as $H_{1DEG}=\int dx \pi j^2/2 uK =\int dx m j^2/2n$, where $m$ is the mass and $n=k_F/\pi$ is the average density. Then, we get $uK=k_F/m=v_F=1$. Incidentally, the resulting current $j=(\rho_R-\rho_L)$ is 
consistent with the chiral charge, Eq.\ref{current-based-chiralApp}.

\section{Current \texorpdfstring{$j$}{j} v.s. Chiral Charge \texorpdfstring{$\rho_{c, sb}$}{rhoc,sb}\label{apx:B}}
The relationship between the current $j$ and chiral charge $\rho_{c, sb}$ naively seems simple, at least when the current $j$ is large enough so that the number of low-speed solitons only is a small portion of the total. However, an extra soliton-antisoliton pair may be produced with zero total momentum, i.e., a soliton appears in the upper right and an antisoliton in the lower left in the energy spectrum (see FIG.\ref{fig:sgEk}). This procedure will not change the total momentum and current, but the number of soliton pairs will increase. Fortunately, it is prohibited due to energy conservation. In this appendix, we will demonstrate that the current $j$ is approximately equal to the chiral charge by considering the charge density profile under a large and instant position-dependent electric field $E(x)$.

After applying a large and instant electric field $E$, the system will contain high-density solitons and antisolitons. In this high-density gas, the $cos$-term can be neglected to describe the behavior of solitons. The Hamiltonian can be written as
\begin{equation}
H=\frac{1}{2}\int dx\left[uK\left(\Pi+\frac{1}{\sqrt{\pi}}eA(x)\right)^2+\frac{u}{K}(\partial_x\Phi)^2\right]
\end{equation}
, where the vector potential is
\begin{equation}
A_1(x)=-ETe^{-x^2/2\sigma^2}\frac{L}{\sqrt{2\pi}\sigma}
\end{equation}
, $E$ is the electric field strength, $T$ is the action time, $\sigma$ is the characteristic length of the vector potential, and $L/\sqrt{2\pi}\sigma$ is a normalization factor. The soliton mass effect can be neglected when $\sigma \ll h/mu^2$. Then, we can use the Heisenberg equation to figure out the charge density profile in the time $t$. The time derivative of the field is
\begin{equation}
\partial_t\Phi=i[H,\Phi]=uK\left(\Pi+\frac{1}{\sqrt{\pi}}eA(x)\right)
\end{equation}
. Then, at the initial time, we have conditions
\begin{equation}
\left<\Phi(x,t=0)\right>=0;\quad \partial\left<\Phi(x,t=0)\right>=\frac{uK}{\sqrt{\pi}}eA(x)
\end{equation}.
On the other hand, we can expand the field $\Phi$ into left and right moving parts, namely,
\begin{equation}
\Phi(x,t)=f(x-ut)+f(x+ut)
\end{equation}.
The function $f(x)$ can be determined by initial conditions and be expressed as
\begin{equation}
f(x)=\frac{\sigma}{2\sqrt{2}}KeET \,erf\left(\frac{x}{\sqrt{2}\sigma}\right) \frac{L}{\sqrt{2\pi}\sigma}
\end{equation}
, where $erf(x)$ is the error function. Hence, the charge density $\rho(x,t)$ and the current density $j(x,t)$ are given by
\begin{widetext}
\begin{eqnarray}
\left<\rho(x,t)\right>&=&\frac{1}{\sqrt{\pi}}\left<\partial_x\Phi\right>=\frac{K}{2\pi}eET\left[e^{-(x-ut)^2/2\sigma^2}-e^{-(x+ut)^2/2\sigma^2}\right]\frac{L}{\sqrt{2\pi}\sigma}\\
\left<j(x,t)\right>&=&-\frac{1}{\sqrt{\pi}}\left<\partial_t\Phi\right>=\frac{uK}{2\pi}eET\left[e^{-(x-ut)^2/2\sigma^2}+e^{-(x+ut)^2/2\sigma^2}\right]\frac{L}{\sqrt{2\pi}\sigma}
\end{eqnarray}
\end{widetext}

. From the above expressions, it is clear that the number $N_{s(\bar{s})}$ of right(left) moving charge wave-packets ((anti)solitons) is $KeETL/2\pi$. After a long time, the solitons and antisolitons will separate in real space and can be easily distinguished by local measurements. 

The average current is 
\begin{equation}
\left<\bar{j}\right>=uKeET/\pi=u(N_s+N_{\bar{s}})/L=u\rho_{c,sb}
\end{equation}
as we expected. Hence, we confirm that the average current $\bar{j}$ and the chiral charge density $\rho_{c, sb}$ are the same, apart from the characteristic speed $u$.

\section{Kinetic Stress-energy Tensor \texorpdfstring{$K^{\mu\nu}$}{K-mu-nu} and its Gauge-invariance\label{apx:A}}

In this appendix, we discuss the stress-energy tensor of a general system with a gauge-invariant and minimal-coupled Lagrangian density $\mathcal{L}=\mathcal{L}(\psi, D_\mu\psi, D_\mu D_\nu\psi,\cdots)$, where $D_\mu=\partial_\mu+ieA_\mu$. Simply, the continuity equation can be derived by applying an infinitesimal translation:
\begin{equation}\label{gen_corr_j5_K}
\partial_\mu T^{\mu1}=\mathcal{F}^1
\end{equation}
, where 
\begin{align}\label{apx:eq:T}
	T^{\mu\nu}=&-\mathcal{L}\eta^{\mu\nu}+\frac{\partial\mathcal{L}}{\partial(\partial_\mu\psi)}\partial^\nu\psi\nonumber\\
	&+\left[\frac{\partial\mathcal{L}}{\partial(\partial_\rho \partial_\mu\psi)}\partial_\rho \partial^\nu\psi+\frac{\partial\mathcal{L}}{\partial(\partial_\mu \partial_\rho\psi)}\partial_\rho \partial^\nu\psi\right.\nonumber\\
	&\left.-\partial_\rho\left(\frac{\partial\mathcal{L}}{\partial(\partial_\rho \partial_\mu\psi)}\partial^\nu\psi\right)\right]+\cdots,\\
	\mathcal{F}^\nu=&-\frac{\partial\mathcal{L}}{\partial A^\mu}\partial^\nu A^\mu-\frac{\partial\mathcal{L}}{\partial (\partial^\rho A^\mu)}\partial^\rho \partial^\nu A^\mu+\cdots.\label{FApp}
	\end{align}
This equation can be modified to the gauge-invariant form, which will be proved in the following, as
\begin{equation}
 \partial_\mu K^{\mu\nu}=ej_\mu F^{\nu\mu},
\end{equation}
(i.e., Eq.\ref{K_conti_F}), where the kinetic stress-energy tensor $K^{\mu\nu}$ is written as:
\begin{equation}{\label{K_gen}}
K^{\mu \nu}=T^{\mu \nu}-ej^\mu A^\nu+C^{\mu \nu}
\end{equation}
, where
\begin{multline}\label{C_munuApp}
	C^{\mu\nu}=\frac{\partial\mathcal{L}}{\partial(\partial_\mu A^\rho)}\partial^\nu A^\rho\\
	+\left[\frac{\partial\mathcal{L}}{\partial(\partial_\mu\partial_\rho A^\rho)}\partial^\nu\partial_\rho A^\rho
	-\partial_\rho\left(\frac{\partial\mathcal{L}}{\partial(\partial_\mu\partial_\rho A^\rho)}\right)\partial^\nu A^\rho\right]\\
	+\cdots.
\end{multline}

In the following, we will show that the kinetic stress-energy tensor $K^{\mu\nu}$ is naturally gauge-invariant. The basic technique is still the Variation method, but in a gauge-invariant approach. Consider the functional expansion of the Lagrangian density $\mathcal{L}$, which is
\begin{equation}\label{20}
\mathcal{L}=\mathcal{L}(\psi(x-a),D_\mu\psi(x-a),D_\mu D_\rho\psi(x-a),\cdots)
\end{equation} 
, around $0$.
The variation of it can be written into three parts:
\begin{align}
\delta \mathcal{L}= &-a_\nu \partial^\nu \mathcal{L}\nonumber\\
&-a_\nu \left[-\frac{\partial\mathcal{L}}{\partial A^\sigma}\partial^\nu A^\sigma-\frac{\partial\mathcal{L}}{\partial (\partial^\rho A^\sigma)}\partial^\nu \partial^\rho A^\sigma+\cdots\right]\nonumber\\
&+\left[-\frac{\partial\mathcal{L}}{\partial(D_\mu\psi)}D_\mu(a^\sigma\partial_\sigma\psi)+\cdots-\text{liner terms of }a_\nu\right]
\end{align}
The first term is the change of the Lagrangian density under the change of the space-time position $x$. The second part is the compensation to the change of the vector potential $A_\mu$ since there is no direct variation of $A_\mu$ in the functional derivative Eq.\ref{20}. The third part is the first and higher derivative terms. The functional derivative of the first and the third term gives us the stress-energy tensor $\partial_\mu T^{\mu\nu}$. The second one is just the $\mathcal{F}^\nu$ (Eq.\ref{FApp}). Hence, we can write the stress-energy tensor as 
\begin{multline}\label{TApp}
T^{\mu\nu}=-\mathcal{L}\eta^{\mu\nu}\\
-\frac{\delta}{\delta (\partial_\mu a_\nu)}\int d^d x\,\left[\mathcal{L}(\psi(x-a),D_\mu\psi(x-a),\cdots)\right.\\
\left.-\text{linear terms of }a_\nu\right].
\end{multline}

In particular, the expression of the stress-energy tensor of 1DEG (Eq.(\ref{S20}),(\ref{S2int})) is 
\begin{eqnarray}{\label{eq:T1deg}}
\begin{aligned}
T_{1DEG}^{00}&=\frac{1}{2m}\bar{D}_1\psi^\dagger D_1\psi\\
&+\left[eA_0-\int dx' V(x-x') \rho(x')  -\mu\right] \rho(x)\\
T_{1DEG}^{10}&=-\frac{1}{2m}\left(\bar{D}_1\psi^\dagger \partial_t\psi+\partial_t\psi^\dagger D_1\psi\right)\\
T_{1DEG}^{01}&=-i\psi^\dagger\partial_x\psi\\
T_{1DEG}^{11}&=i\psi^\dagger D_0\psi\\
&+\frac{1}{2m}\left(\bar{D}_1\psi^\dagger \partial_x\psi+\partial_x\psi^\dagger D_1\psi-\bar{D}_1\psi^\dagger D_1\psi\right)\\
&+\left[\int dx' V(x-x') \rho(x')  +\mu\right] \rho(x)
\end{aligned} 
\end{eqnarray}
, regardless of the Lagrangian density is non-local.

To compare the expression of the tensor $C^{\mu\nu}$ (Eq.\ref{C_munuApp}) with the one of the stress-energy tensor $T^{\mu\nu}$ (Eq.\ref{TApp}), we also can view the tensor $C^{\mu\nu}$ as a functional derivative:
\begin{multline}
C^{\mu\nu}=-\frac{\delta}{\delta (\partial_\mu a_\nu)}\int d^d x\,\left[\mathcal{L}(\psi(x),D^-_\mu\psi(x),\cdots)\right.\\\left.-\text{linear terms of }a_\nu\right].
\end{multline}
, where $D^-_\mu=D_\mu -ie a_\nu\partial^\nu A_\mu$.

The functional derivative form of the operator $-j^\mu A^\nu$ can be derived from the definition of the current operator, i.e., $j^\mu=-\delta L/\delta A_\mu$. It is
\begin{multline}
-j^{\mu}A^{\nu}=-\frac{\delta}{\delta(\partial_\mu a_\nu)}\int d^d x\,\left[\mathcal{L}(\psi(x),D^c_\mu\psi(x),\cdots) \right.\\\left.-\text{liner terms of }a_\nu\right].
\end{multline}
, where $D^c_\mu=D_\mu -ieA^\nu \partial_\mu a_\nu$. It is easy to check by using the chain rule of the functional derivative, i.e.,
\begin{align}
\frac{\delta}{\delta(\partial_\mu a_\nu)}&=\int d^d x'\,\frac{\delta}{\delta(A^\rho\partial_\sigma a_\rho)}\frac{\delta(A^\rho\partial_\sigma a_\rho)}{\delta(\partial_\mu a_\nu)}\nonumber\\
&=\frac{\delta}{\delta(A^\rho\partial_\mu a_\rho)}A^\nu.
\end{align}

Then, the kinetic stress-energy tensor is finally written as
\begin{multline}
K^{\mu\nu}=-\mathcal{L}\eta^{\mu\nu}-\frac{\delta}{\delta(\partial_\mu a_\nu)}\int d^d x\,\left[\mathcal{L}(\psi',D'_\mu\psi',\cdots) \right.\\\left.-\text{liner terms of }a_\nu\right].
\end{multline}
, where $\psi'=\psi -a_\nu \partial^\nu \psi$ and $D'_\mu=D_\mu -ie(a_\nu \partial^\nu A_\mu+A^\nu\partial_\mu a_\nu)$. The high-order terms of $a_\nu$ are neglected in the above expression since they do not contribute.

Since the Lagrangian density is gauge-invariant, we can perform a gauge transformation with the argument $-iea_\nu A^\nu$. Therefore, the $K^{\mu\nu}$ is also equal to
\begin{multline}
K^{\mu\nu}=-\mathcal{L}\eta^{\mu\nu}-\frac{\delta}{\delta(\partial_\mu a_\nu)}\int d^d x\,\left[\mathcal{L}(\psi'',D''_\mu\psi'',\cdots)\right.\\
\left.-\text{liner terms of }a_\nu\right]
\end{multline}
, where $\psi''=\psi -a_\nu D^\nu \psi$ and $D''_\mu=D_\mu-iea^\nu F_{\nu\mu}$. The linear term here is just $a_\nu \partial^\nu \mathcal{L}$ since it all comes from the change of the position $x$. We notice that
the Lagrangian density above in the expression of $K^{\mu\nu}$ is gauge-invariant under any gauge transformations since the covariant derivative $D_\mu$ and the field strength tensor $F_{\mu\nu}$ have been used in the variation of the field $\psi$ and the vector potential $A_\mu$ instead of partial derivatives. Together with the linear terms' gauge invariance, we can conclude that the kinetic stress-energy tensor is naturally gauge-invariant. The concrete expression for the $K^{\mu\nu}$ up to the second order is 
\begin{align}\label{apx:eq:K}
K^{\mu\nu}=&-\mathcal{L}\eta^{\mu\nu}+\frac{\partial\mathcal{L}}{\partial(D_\mu\psi)}D^\nu\psi\nonumber\\
&+\left\{-ie\frac{\partial\mathcal{L}}{\partial(D^\rho D_\mu\psi)}F^{\nu\rho}\psi\right.\nonumber\\
&+\left[\frac{\partial\mathcal{L}}{\partial(D_\rho D_\mu\psi)}D_\rho D^\nu\psi+\frac{\partial\mathcal{L}}{\partial(D_\mu D_\rho\psi)}D_\rho D^\nu\psi\right.\nonumber\\
&\left.\left.-\partial_\rho\left(\frac{\partial\mathcal{L}}{\partial(D_\rho D_\mu\psi)}D^\nu\psi\right)\right]\right\}+\cdots.
\end{align}

Incidentally, the expression for the $K^{01}$ term in the generalized 1DEGs can be simplified due to the presence of only one first $t$-derivative term in the Lagrangian. Consequently, only the second term on the left side of Eq.\ref{apx:eq:K} remains non-zero, leading to the simplified form:
\begin{equation}
 K^{01}=-i\psi^\dagger D_1\psi.
\end{equation}

\section{\texorpdfstring{$\boldsymbol{j}_{\rm{CME}}$}{jCME} in the Chiral Kinetic Theory\label{apx:D}}
In this appendix, we will document the procedures for deriving the chiral magnetic effect(CME), i.e. Eq.\ref{eq:jcme}, in the chiral kinetic theory. 

First, the total current in the chiral kinetic theory is obvious to find by definition\cite{Stephanov2012}:
\begin{align}
\boldsymbol{j}&=\int_{\boldsymbol k} \sqrt{G_k}f\dot{ \boldsymbol{r}}\nonumber\\
&=\int_{\boldsymbol k} f \boldsymbol{v_k}+ e\boldsymbol{E}\times \int_{\boldsymbol k} f \boldsymbol{\Omega}_{ \boldsymbol{k}}+e\boldsymbol{B}\int_{\boldsymbol k} f \left(\boldsymbol{v_k}\cdot \boldsymbol{\Omega}_{ \boldsymbol{k}}\right).\label{jcmeApp}
\end{align}
Here the first term in Eq.\ref{jcmeApp} is the regular current; the second term is the anomalous Hall current; the third term which is along the magnetic field direction is the chiral magnetic current. Therefore, we obtain a comprehensive equation for the chiral magnetic effect (CME) current as follows:
\begin{equation} \boldsymbol{j}_{\rm{CME}}=e^2\boldsymbol{B}\int_{\boldsymbol k} f \left(\partial_{\boldsymbol k} \varepsilon(\boldsymbol{k})\cdot \boldsymbol{\Omega}_{ \boldsymbol{k}}\right).
\end{equation}

Then, we use the integral by parts and determine that
\begin{align}
\int_{\boldsymbol k} f \left(\partial_{\boldsymbol k} \varepsilon(\boldsymbol{k})\cdot \boldsymbol{\Omega}_{ \boldsymbol{k}}\right)&= \int_{\boldsymbol k} \nabla_{\boldsymbol k}\cdot\left(\varepsilon f \boldsymbol{\Omega}_{ \boldsymbol{k}}\right) -\varepsilon \nabla_{\boldsymbol k}\cdot\left(f \boldsymbol{\Omega}_{ \boldsymbol{k}}\right).
\end{align}
The first term is a boundary term and thus vanishes. The second term can be written as
\begin{align}\label{d4App}
-\int_{\boldsymbol k} \varepsilon \nabla_{\boldsymbol k}\cdot\left(f \boldsymbol{\Omega}_{ \boldsymbol{k}}\right)=- \int_{\boldsymbol k} \varepsilon \left(\nabla_{\boldsymbol k}f\right)\cdot\boldsymbol{\Omega}_{ \boldsymbol{k}}- \varepsilon f\left(\nabla_{\boldsymbol k}\cdot\boldsymbol{\Omega}_{ \boldsymbol{k}}\right).
\end{align}
At zero temperature, the first term can be expressed as a Fermi-surface integral given that $\nabla_{\boldsymbol k}f$ is only non-zero on Fermi surfaces:
\begin{equation}
-\int_{\boldsymbol k} \varepsilon \left(\nabla_{\boldsymbol k}f\right)\cdot\boldsymbol{\Omega}_{ \boldsymbol{k}}=\frac{1}{(2\pi)^3}\sum_\alpha \int_{S_\alpha} d^2 \mathcal{F}\,\varepsilon.
\end{equation}
The second term in Eq.\ref{d4App} is non-zero due to the term $\nabla_{\boldsymbol k}\cdot\boldsymbol{\Omega}_{ \boldsymbol{k}}$ at two Weyl points. However, it will not contribute to the CME current. The reason is that the CME current(Eq.\ref{jcmeApp}) is the sum of all bands. When we consider the fully occupied lower Weyl band, since $\nabla_{\boldsymbol k}f$ is always zero, the first term in Eq.\ref{d4App} vanishes. The second term for the lower Weyl band is identical to the upper Weyl band except for the opposite sign. Therefore, when accounting for the contributions of all bands to the CME current, the second term in Eq. \ref{d4App} will cancel out, leaving only the first term from the upper Weyl band.

Finally, we can express the CME current as
\begin{equation}\label{jcmefAPP} \boldsymbol{j}_{\rm{CME}}=\frac{e^2\boldsymbol{B}}{4\pi^2}\Delta \mu,
\end{equation}
where
\begin{align}
&\Delta \mu=\frac{1}{2\pi}\sum_\alpha \int_{S_\alpha} d^2 \mathcal{F}\,\varepsilon(\boldsymbol{k}).
\end{align}
Here Eq.\ref{jcmefAPP} is a general expression in the zero-temperature limit. If we assume that the electron distribution functions are in equilibrium around each Weyl node, i.e. $\varepsilon(\boldsymbol k)=const.$ for $\boldsymbol{k} \in S_\alpha$, we can define the chiral chemical potential, i.e. $\mu_\alpha$, and find that $\Delta \mu$ is exactly the chiral chemical potential difference, $\mu_R-\mu_L$. Then, Eq.\ref{jcmefAPP} is consistent with references\cite{Armitage2018,Son2013}. 

As an illustration, it is straightforward to derive the expression for a simple Weyl semi-material using the energy dispersion $\varepsilon_\alpha(\boldsymbol{k})=\gamma_\alpha vk$ and the Berry curvature $\boldsymbol{\Omega}{\boldsymbol{k}\alpha}=\gamma\alpha\hat{\boldsymbol{k}}/2k^2$ around the Weyl nodes $\gamma_\alpha$ \cite{Stephanov2012}. The obtained result evidently corresponds to Eq. \ref{jcmefAPP}.

\subsection{Berry Curvature in Fermi liquids}
While the topological Fermi liquid~\cite{Haldane2004} as well as the chiral kinetic theory discussed above is defined in terms of Berry curvature on the Fermi surface, Berry curvatures are typically calculated from non-interacting band structures of Fermions. This leads to a question of how to precisely define Berry curvature on a 
strongly interacting Fermi surface.

Basically, the Berry curvature is defined for non-interacting systems. Shou-cheng Zhang defined invariants for interacting systems with Green functions.\cite{Wang2012} The density matrix for quasi-particles at the momentum $\boldsymbol{k}$ is defined by $\rho_{\boldsymbol k}=N_{\boldsymbol k}\int_{\omega\sim \epsilon_{\boldsymbol k}} d\omega A_{\boldsymbol k}(\omega)$, where $A_{\boldsymbol k}(\omega)=\rm{Im} [G_{\boldsymbol k}(\omega)]$ is the spectral function matrix. $N_{\boldsymbol k}$ is a normalization that ensures that $\rho_{\boldsymbol k}^2=\rho_{\boldsymbol k}$ and $Tr[\rho_{\boldsymbol k}]=1$ for $\boldsymbol{k}$ on the Fermi surface. The density matrix near a non-degenerate Fermi surface can be expanded in terms of a wave-function $\rho_{\boldsymbol k}({\boldsymbol r},{\boldsymbol r'})=u_{\boldsymbol k}({\boldsymbol r})u_{\boldsymbol k}^*({\boldsymbol r'})$ due to the Landau liquid property. This shows that the product
\begin{equation}
Tr[\rho_{\boldsymbol k}\rho_{{\boldsymbol k}_1}\rho_{{\boldsymbol k}_2}]=\langle u_{\boldsymbol k}|u_{{\boldsymbol k}_1}\rangle\langle u_{{\boldsymbol k}_1}|u_{{\boldsymbol k}_2}\rangle\langle u_{{\boldsymbol k}_2}|u_{{\boldsymbol k}}\rangle=e^{i\Omega_{\boldsymbol k} \mathcal{A}},
\end{equation}
where $\mathcal{A}$ is the area of the triangle. Expanding $\rho_{{\boldsymbol k}_j}\simeq\rho_{\boldsymbol k}+({\boldsymbol k}_j-{\boldsymbol k})\cdot\partial_{\boldsymbol k}\rho_{\boldsymbol k}$, and we can find the Berry curvature to be $\Omega_{\boldsymbol k}=Tr[\partial_{\boldsymbol k}\rho\times (\rho\partial_{\boldsymbol k}\rho)]$. It means the berry curvature on the Fermi surface can be exactly defined by using Green's functions.

\section{Second-order Term of the ac Hall Conductivity in the Chern Insulator\label{apx:E}}
In this part, rather than directly using the Kubo formula to calculate the ac hall conductivity in Weyl, we will estimate it in the Chern Insulator(CI). Since the Weyl systems can be described as a stack of the Chern insulator, the integral of ac Hall conductivity in CI over one parameter $k_z$ is just the one in Weyl. However, if we only need to prove that the ac Hall conductivity in Weyl is non-zero and non-universal, the result of the ac Hall conductivity in CI will be sufficient.

The general Kubo formula for the ac conductivity is 
\begin{equation}
\sigma_{xy}(\omega)=\frac{1}{i\omega}\sum_{n\neq0}\left[\frac{\left<0|J_y|n\right>\left<n|J_x|0\right>}{\omega+(E_n-E_0)}-\frac{\left<0|J_x|n\right>\left<n|J_y|0\right>}{\omega-(E_n-E_0)}\right]
\end{equation}
, where $J_i$ is the current operator, the state $0$ and $n$ represents the many-body ground state and the $n$-th excited state. To expand the above expression in series of $\omega$, we have
\begin{equation}
\begin{aligned}
\sigma_{xy}(\omega)=&-\frac{i}{\omega}\sum_{n\neq0}\frac{\left<0|J_y|n\right>\left<n|J_x|0\right>+\left<0|J_x|n\right>\left<n|J_y|0\right>}{E_n-E_0}\\
&+i\sum_{n\neq0}\frac{\left<0|J_y|n\right>\left<n|J_x|0\right>-\left<0|J_x|n\right>\left<n|J_y|0\right>}{(E_n-E_0)^2}\\
&-i\omega\sum_{n\neq0}\frac{\left<0|J_y|n\right>\left<n|J_x|0\right>+\left<0|J_x|n\right>\left<n|J_y|0\right>}{(E_n-E_0)^3}\\
&+i\omega^2\sum_{n\neq0}\frac{\left<0|J_y|n\right>\left<n|J_x|0\right>-\left<0|J_x|n\right>\left<n|J_y|0\right>}{(E_n-E_0)^4}\\
&+\cdots
\end{aligned}
\end{equation}

The first term and all odd order terms vanish, because of gauge-invariance. Alternatively, it can quickly be seen by rotation-invariance which ensures that the expression should be invariant under $x \rightarrow y$ and $y \rightarrow - x$.

For a simple Chern insulator, the single particle Hamiltonian in the momentum space as 
\begin{equation}
H(\boldsymbol{k})=\vec{d}(\boldsymbol{k})\cdot \vec{\sigma}
\end{equation}
with
\begin{equation}
\begin{aligned}
d_1(\boldsymbol{k})&=\sin(k_x);\\
d_2(\boldsymbol{k})&=\sin(k_y);\\
d_3(\boldsymbol{k})&=2-m-\cos(k_x)-\cos(k_y)
\end{aligned}
\end{equation}
, where the $\vec{\sigma}$ is the Pauli matrices. After simplifying, we can write the DC conductivity as 
\begin{equation}
\sigma^{DC}_{xy}=\frac{e^2}{(2\pi)^2}\int_{\boldsymbol{T}_2}d^2 k\, \mathcal{F}_{xy}
\end{equation}
and the second-order conductivity as
\begin{equation}
\sigma^{(2)}_{xy}=\omega^2\frac{e^2}{(2\pi)^2}\int_{\boldsymbol{T}_2}d^2 k\, \frac{\mathcal{F}_{xy}}{(E_+-E_-)^2}
\end{equation}
with the Berry curvature
\begin{equation}
\mathcal{F}_{xy}=\frac{\partial\mathcal{A}_x}{\partial k_y}-\frac{\partial\mathcal{A}_y}{\partial k_x}
\end{equation}
and the Berry connection
\begin{equation}
\mathcal{A}_i=i\left<-,k\right|\frac{\partial}{\partial k_i}\left|-,k\right>. 
\end{equation}
, where the state $\left|-,k\right>$ is the lower energy eigenstate. The high-order conductivity can be computed in the same sense\cite{Tewari_2008,Pershoguba_2022}.
Actually, the integral in the DC conductivity is related to the Chern number, which is an integer universally. This is just the famous quantized Hall conductivity of the Chern insulator. However, the second-order conductivity cannot be related to the Chern number. In the following, we will show the numerical result of this second-order conductivity.
\begin{figure}
\centering
\includegraphics[width=0.5\textwidth]{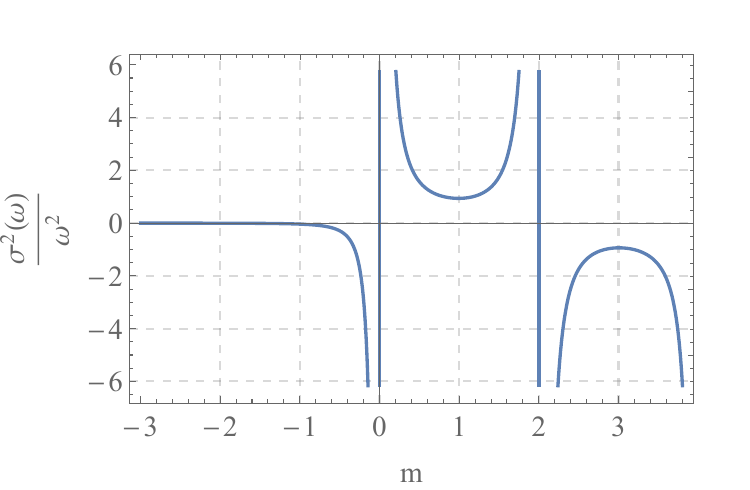}\label{fig.2c}
\caption{The second order conductivity $\sigma_{xy}^{(2)}$ v.s. the parameter $m$. It is not quantized and non-universal}
\end{figure}
Based on the Hamiltonian of the simple Chern insulator, the Berry curvature can be analytically computed and be written as
\begin{equation}
\mathcal{F}_{\mu\nu}=-\frac{1}{2}\epsilon_{\alpha\beta\gamma}\hat{d}_\alpha\partial_{k_\mu}\hat{d}_\beta\partial_{k_\nu}\hat{d}_\gamma
\end{equation}
with
\begin{equation}
\hat{d}_\alpha(\boldsymbol{k})=d_\alpha(\boldsymbol{k})/d(\boldsymbol{k}).
\end{equation}
Hence, we have the expression for $\mathcal{F}_{xy}$,
\begin{widetext}
\begin{equation}
\mathcal{F}_{xy}=\frac{\cos k_y + 
\cos k_x [1 + (-2 + m) \cos k_y]}{2 [(-2 + m + \cos k_x + \cos k_y)^2 + 
\sin^2 k_x + \sin^2 k_y]^{3/2}}.
\end{equation}
\end{widetext}

The numerical results of the second-order conductivity are shown in Fig.\ref{fig.2c}. The second-order conductivity is not zero or quantized. It is non-universal.

\end{document}